\documentclass[%
aip,
amsmath,amssymb,
author-numerical,%
]{revtex4-2}

\def \bF {\pmb{F}}
\def \bH {\pmb{H}}

\def \bx {\pmb{x}}

\def \bw {\pmb{w}}
\def \bs {\pmb{s}}
\def \by {\pmb{y}}

\usepackage{graphicx}
\usepackage{dcolumn}
\usepackage{bm}
\usepackage[dvipsnames]{xcolor}
\usepackage[normalem]{ulem}
\usepackage[title]{appendix}
\usepackage{comment}
\usepackage{float}
\usepackage{mathtools}

\usepackage{ subfigure }

\usepackage{hyperref}
\usepackage{multirow}
\usepackage[utf8]{inputenc}
\usepackage{amsthm}

\newtheorem{definition}{Definition}
\newtheorem{remark}{Remark}

\newcommand\itema{\item[\textbf{A1:}]}
\newcommand\itemb{\item[\textbf{A2:}]}
\newcommand\itemc{\item[\textbf{A3:}]}

\usepackage{epstopdf}

\usepackage[shortlabels]{enumitem}
\graphicspath{{./figures/}}
\theoremstyle{definition}

\newcolumntype{C}[1]{>{\centering\arraybackslash$}m{#1}<{$}}
\newlength{\mycolwd}                                         
\usepackage{pgfplots}
\begin{document}
		
			\title{\bf {Matryoshka and Disjoint Cluster Synchronization of Networks}}
	\author{Amirhossein Nazerian}
	\affiliation{Mechanical Engineering Department, University of New Mexico, Albuquerque, NM, 87131}
	\author{Shirin Panahi}
	\affiliation{Mechanical Engineering Department, University of New Mexico, Albuquerque, NM, 87131}
	\author{Ian Leifer}
	\affiliation{Levich Institute and Physics Department, City College of New York, New York, NY, 10031}
	\author{David Phillips}
	\affiliation{Mathematics Department, United States Naval Academy, Annapolis, MD 21401}
	\author{Hernán A. Makse}
	\affiliation{Levich Institute and Physics Department, City College of New York, New York, NY, 10031}
	\author{Francesco Sorrentino}
	\email{fsorrent@unm.edu}
	\affiliation{Mechanical Engineering Department, University of New Mexico, Albuquerque, NM, 87131}		
		
	\begin{abstract}
The main motivation for this paper is to present a definition of network synchronizability for the case of cluster synchronization \textcolor{black}{(CS)}, in an analogous fashion to Barahona and Pecora\cite{barahona:2002} for the case of complete synchronization. We find this problem to be substantially more complex than the original one. We distinguish between the two cases of networks with intertwined clusters and no intertwined clusters and between {the two cases that the master stability function is negative either in a bounded range or in an unbounded range of its argument}. \textcolor{black}{We first obtain a definition of synchronizability that applies to each individual cluster within a network and then attempt to generalize this definition to the entire network.}  For CS, \textcolor{black}{the synchronous solution of} each cluster may be stable independent of the \textcolor{black}{stability of the other clusters}, which \textcolor{black}{results in possibly different ranges in which each cluster synchronizes (isolated CS.)} For each pair of clusters, we distinguish between three different cases: Matryoshka Cluster Synchronization (when the range of the stability \textcolor{black}{of the synchronous solution} for one cluster is included in that of the other cluster), Partially Disjoint Cluster Synchronization (when the ranges of stability \textcolor{black}{of the synchronous solutions} partially overlap), and Complete Disjoint Cluster Synchronization (when the ranges of stability \textcolor{black}{of the synchronous solutions} do not overlap.)

\bigskip

	\end{abstract}

	\maketitle

\textbf{Previous work in the area of cluster synchronization of networks has shown that is possible for different clusters of nodes to synchronize in different ranges of the coupling strength. We  define the cluster synchronizability as the range of the coupling strength in which \textcolor{black}{all clusters have locally stable synchronous solutions}. We distinguish between three different types of cluster synchronization, which we call Matryoshka, Partially Disjoint, and Complete Disjoint, and discuss their role in affecting cluster synchronizability for both the cases of synthetic and real-world networks.}

\section{Introduction} \label{s:intro}

 Cluster synchronization (CS) in networks of coupled oscillators
has been the subject of vast research efforts, see e.g., \cite{belykh2008cluster,NSG,dahms2012cluster,fu2014synchronization,kanter2011nonlocal,rosin2013control,williams2013experimental,nicosia2013remote,schaub2016, Panahi2021Group}. \textcolor{black}{CS often occurs in networks with connectivity  described by an adjacency matrix, but it can also occur in networks with connectivity described by a Laplacian matrix, with sums of the entries in all the rows equal to zero. Reference \cite{KANEKO1990137} studied the emergence of CS in networks of globally coupled chaotic maps and described the emergence of multi-stability of CS patterns in such networks. References \cite{Belykh2000Hierarchy, Belykh2003Persistent, BELYKH2003CLUSTER} studied CS in 1D, 2D, and 3D lattices where the network topology is described by a Laplacian matrix. These papers also showed that the synchronization manifold corresponding to a given pattern of CS can be embedded in the synchronization manifold corresponding to another pattern of CS, which is consistent with the emergence of different patterns of CS in these networks. }

 \textcolor{black}{Recent work has discussed the existence of clusters of nodes in both undirected \cite{Morone19} and directed \cite{MoronePNAS20, LeiferBMC20} biological networks \cite{golubitsky2006nonlinear}. Here we focus on synchronizability, which is concerned with the stability of the synchronous solution.} There is a known definition of network synchronizability \cite{barahona:2002,huang:2009}, which specifically applies to the case of complete synchronization and describes how `synchronizable' a given network is, regardless of the particular dynamics of the network nodes and the specific choice of the node-to-node interaction. The synchronizability measures the range of the coupling strength (hereafter, $\sigma$) over which the synchronous solution is stable. Despite the broad interest in CS and its relevance to biological networks, no definition of cluster synchronizability, which is related to the stability of CS, has been given in the literature. The goal of this paper is to provide such definition.

\textcolor{black}{To motivate the need for introducing a definition of cluster synchronizability, consider the following questions: (1) Does either addition or removal of some of network links increase/decrease the range of a parameter that supports stability of CS? (2) Which one of a set of networks supports emergence of CS in a larger range of a relevant parameter? In what follows}
we will see that the case of CS is considerably more complex than the case of complete synchronization and a variety of alternative scenarios may occur, which we call Matryoshka synchronization, Partially Disjoint synchronization, and Completely Disjoint synchronization; yet, under appropriate assumptions, we will derive a definition of cluster synchronizability and use it to compare several networks.

In Sec.\ \ref{CompSynch}, we briefly review the case of complete synchronization for networks described by the Laplacian matrix, and in Sec.\ \ref{sec:clust}, we study the case of cluster synchronization for networks described by the adjacency matrix. In both cases, we focus on stability of the synchronous solution corresponding to the network minimum balanced coloring \cite{belykh2011mesoscale}, i.e., the solution for which the fewest sets of nodes for that network achieve synchronization. In Sec.\,\ref{sec:real} we investigate the case of real network topologies.


\section{The Case of Complete Synchronization} \label{CompSynch}
First, we briefly review the definition of synchronizability for the case of complete synchronization. Consider a network of coupled dynamical systems that evolve based on the following set of equations,
\begin{equation} \label{net:Lclust}
    \dot{\bx}_i(t)=\bF({\bx}_i(t))-\sigma \sum_j L_{ij} \bH(\bx_j(t)),
\end{equation}
for $i=1,...,N$, where ${\bx}_i \in \mathbb{R}^m$ is the state of node $i$, $\bF : \mathbb{R}^m \rightarrow \mathbb{R}^m$, describes the dynamics of each individual uncoupled system, the function $\bH : \mathbb{R}^m \rightarrow \mathbb{R}^m$ describes how two systems interact with one another, $\sigma$ is a scalar parameter that measures the strength of the coupling.
The symmetric Laplacian matrix $L=\{L_{ij}\}$ is defined as $L = D - A$ where the symmetric adjacency matrix $A=\{A_{ij}\}$ describes the network connectivity, i.e., $A_{ij}=A_{ji} \neq 0$ ($A_{ij}=A_{ji} = 0$) if node $j$ is (not) connected to node $i$ and vice versa, and the diagonal matrix $D$ has diagonal entries that are equal to the degree of each node, $D_{ii} = \sum_j A_{ij}$. The Laplacian matrix is positive definite (see Sec.\,I of the Supplementary Information.) Also, since $L$ has sum over its rows equaling zero, the eigenvalues of $L$ are
$ 0 = \lambda_1 \leq \lambda_2 \leq ... \leq \lambda_N$ \cite{fiedler1973algebraic,chung1997spectral}. The Laplacian eigenvalue $\lambda_2$ is known as the algebraic connectivity \cite{fiedler1973algebraic,mohar1992laplace,merris1994laplacian}.

Based on the functions $\bF$ and $\bH$, the network can be synchronized either in a bounded interval or in an unbounded interval of the coupling strength $\sigma$ (see Sec.\ I of the Supplementary Information.) The network synchronizability measures the range of the coupling strength $\sigma$ over which the synchronous state is stable. According to whether this range is bounded or unbounded, there are two alternative definitions of network synchronizability. 

\begin{definition} \textbf{Network synchronizability for the case that the $\sigma$-interval is bounded.} \label{ComNetBounded}
In the case of synchronization within a bounded interval of the coupling strength $\sigma$, the network synchronizability is the eigenratio,
\begin{equation} \label{ratio}
  \textcolor{black}{\alpha_{\textnormal{b}}}  = \frac{\lambda_2}{\lambda_N} \leq 1. 
\end{equation}
The larger the eigenratio $\textcolor{black}{\alpha_{\textnormal{b}}}$, the more synchronizable the network.
\end{definition}
Based on this definition, the best possible synchronizability is obtained for $\lambda_2= \lambda_3=...=\lambda_N$, i.e., all equal eigenvalues except for $\lambda_1=0$. 
We note that the original eigenratio introduced in \cite{barahona:2002} is equal to $1/\textcolor{black}{\alpha_{\textnormal{b}}}=\lambda_N/\lambda_2 \geq 1$ (the smaller this other eigenratio, the more synchronizable the network.)


\begin{definition} \textbf{Network synchronizability for the case that the $\sigma$-interval is unbounded.} \label{ComNetUnbounded}
In the case of synchronization within an unbounded interval of the coupling strength, $\sigma$, the network synchronizability is defined in terms of the second smallest eigenvalue $\lambda_2$ of the network Laplacian,
\begin{equation}
   \textcolor{black}{\alpha_{\textnormal{u}}} = \lambda_2.
\end{equation}
The larger $\textcolor{black}{\alpha_{\textnormal{u}}}$, the more synchronizable the network. 
\end{definition}
Based on the choice of the functions $\bF$ and $\bH$, either Def.\ \ref{ComNetBounded} or Def.\ \ref{ComNetUnbounded} applies. Both definitions are solely a function of the network topology, which in either case, enables a direct comparison between different networks in terms of their synchronizability, see e.g. \cite{barahona:2002,Ni:Mo,bocc1,bocc2,sorrentino2007synchronizability}.

\section{The Case of Cluster Synchronization} \label{sec:clust}
Consider a network with dynamical equations,
\begin{equation} \label{net:Aclust}
    \dot{\bx}_i(t)=\bF({\bx}_i(t))+\sigma \sum_{j = 1}^{N} \tilde{A}_{ij} \bH(\bx_j(t)),
\end{equation}
where $i = 1, ..., N$. The symmetric matrix $\tilde{A} = \left(A - \kappa I_N \right)$ is the shifted version of the symmetric adjacency matrix $A$, where $I_N$ is the $N$-dimensional identity matrix, and the scalar $\kappa \in \mathbb{R}$. Since the rows of the matrix $\tilde{A}$ do not typically sum to a constant, it is not possible for all the network nodes to synchronize on the same time evolution. Instead, it is possible for sets of nodes to synchronize in clusters. \textcolor{black}{The matrix $\tilde{A}=\{\tilde{A}_{ij}\}$ from Eq.\ \eqref{net:Aclust} describes the  network topology. This network can also be represented as a set of nodes, $\mathcal{V} = \left\{ v_1, \hdots, v_N \right\}$, where the dynamical state of the node $v_i$ is given by $\bx_i(t)$ in Eq. \eqref{net:Aclust}. 
\begin{definition} \textbf{Equitable Clusters (Balanced Coloring).} \label{ecp}
Consider the network defined by the shifted adjacency matrix, $\tilde{A}$. The set of the network nodes $\mathcal{V}$ can be partitioned into subsets of nodes called equitable clusters, 
$\mathcal{C}_1,\mathcal{C}_2,..,\mathcal{C}_C$, $\cup_{i=1}^C \mathcal{C}_i=\mathcal{V}$, $\mathcal{C}_i \cap \mathcal{C}_j=\emptyset$ for $i \neq j$, where 
\begin{equation}
  \sum_{h \in \mathcal{C}_{\ell}} \tilde{A}_{ih} = \sum_{h \in \mathcal{C}_{\ell}} \tilde{A}_{jh}, \quad \begin{aligned} \forall i,j \in \mathcal{C}_k \\ \forall \mathcal{C}_k,\mathcal{C}_{\ell} \subset \mathcal{V}. \end{aligned}
\end{equation}
The number of nodes in cluster $\mathcal{C}_k$ is $n_k$, $\sum_{k=1}^C n_k=N.$ A partition of the set of the network nodes $\mathcal V$ into equitable clusters is also called a balanced coloring.
\end{definition}
\begin{definition} \textbf{Minimum Balanced Coloring.} 
A minimum balanced coloring is a partition of the set of the network nodes $\mathcal V$ into equitable clusters 
$\mathcal{C}_1,\mathcal{C}_2,..,\mathcal{C}_K$ with the minimum number $K$ of clusters. A fast algorithm for computing the network minimum balanced coloring  from knowledge of the matrix $\tilde{A}$ was proposed in \cite{belykh2011mesoscale}. 
\end{definition}
A trivial (non-trivial) cluster is one with only one node (more than one node) in it. We call $\tilde K \leq K$ the number of nontrivial clusters. Without loss of generality, we order the clusters such that the first $k=1,...,\tilde K$ clusters are nontrivial and the remaining $k={\tilde K}+1,...,K$ clusters are trivial.}

The shifting parameter, $\kappa$ is used to aid in our network analysis and has a natural interpretation. We note here that in the case of complete synchronization, Eq.\ \eqref{net:Lclust}, the Laplacian matrix has eigenvalues all with the same sign, while the adjacency matrix $A$ has typically both positive and negative eigenvalues, which provides the motivation for introducing the shift $-\kappa.$ 
\textcolor{black}{The presence of autoregulation loops, in the form of inhibitory interactions, has been documented in bacterial transcriptional regulatory networks, especially in the E. coli \cite{alon-motif}, and been associated with a better dynamical response of these networks\cite{purcell2010comparative}. It has been reported that these autoregulation loops are common motifs observed in biological networks \cite{alon-motif,alon-ecoli,alon}.}
 \textcolor{black}{Alternatively, by rewriting  Eq.\  \eqref{net:Aclust} as 
 $
    \dot{\bx}_i(t)=\tilde{\bF}({\bx}_i(t))+\sigma \sum_{j = 1}^{N} {A}_{ij} \bH(\bx_j(t)),$
where $\tilde{\bF}({\bx}_i)=\bF({\bx}_i)-\kappa \bH({\bx}_i) $, the effect of the shift $-\kappa$ can be seen as  incorporated in the dynamics of the individual nodes, without affecting the network connectivity.  Therefore, the shift $-\kappa$ can have two different interpretations, either as a modifier of the network topology by introducing an additional autoregulation loop on each node or as a modifier of the individual node dynamics. Regardless, it is important to emphasize that application of the shift $-\kappa$ is \textit{de facto} needed in most cases for the emergence of stable CS and is in fact  commonly assumed throughout the literature, see e.g. \cite{pecora:2014cluster}.
Therefore, in what follows we will proceed under the assumption that a suitable value of $\kappa$ is given (see our assumption A1 that follows), and study stability of CS as a function of the other parameter $\sigma$.}
In Sec.\,III of the Supplementary Information, we will discuss the application of the theory to the case of negative $\kappa$, but for now, we assume that $\kappa$ is positive.


{The network allows a flow-invariant cluster synchronization solution $\{{\bs}_1(t),...,{\bs}_K(t)\}$, such that
$\bx_i(t)=\bs_k (t)$, $i \in \mathcal{C}_k$, $k=1,...K$ \cite{golubitsky2006nonlinear}. The set of equalities $\bx_i(t)=\bx_j(t)$, $\forall i,j \in \mathcal{C}_k$  defines the invariant CS manifold. By setting $\bx_i(t)=\bs_k (t)$, Eq.\ \eqref{net:Aclust} can be rewritten,
\begin{equation} \label{net:Q}
    \dot{\bs}_k(t)=\bF({\bs}_k(t))+\sigma \sum_j Q_{kj} \bH(\bs_j(t)),
\end{equation}
$k=1,...,K$,
where the $K$-dimensional quotient matrix $Q=\{Q_{kj}\}$ satisfies
\begin{equation}
\label{mat:Q}
    Q=(Z^T Z)^{-1} Z^T A Z = Z^\dagger A Z,
\end{equation}
and the $N \times K$-dimensional indicator matrix $Z=\{Z_{ij}\}$ is such that $Z_{ij}=1$ if node $i$ belongs to cluster $\mathcal{C}_j$ and $Z_{ij}=0$ otherwise, $j=1,...,K$. $Z^\dagger$ indicates the Moore–Penrose inverse of the matrix $Z$.}

\textcolor{black}{
\begin{remark}
There can be multiple sets of equitable clusters associated with a given adjacency matrix $A$, see e.g. \cite{kamei2013computation}. Multiple sets of equitable clusters (in addition to the case in which all the nodes are in only one cluster) are also possible in networks with connectivity described by a Laplacian network, with dynamics given by Eq.\ \eqref{net:Lclust} \cite{SA}. The study of stability of the  corresponding CS solutions can be performed using the approaches of \cite{Belykh2000Hierarchy, Belykh2003Persistent, BELYKH2003CLUSTER, SA,siddique2018symmetry,Zhang2020,Panahi2021Group}. For simplicity, in what follows we focus on stability of the one CS solution associated with the minimum balanced coloring (Definition 3), but our work is directly generalizable to any CS solution associated with an equitable cluster partition for either Eq.\ \eqref{net:Lclust} or Eq.\ \eqref{net:Aclust}. 
\end{remark}
In the rest of this paper, we will focus on the case that the number of clusters $1<K<N$, which excludes the cases $K=1$ corresponding to complete synchronization and $K=N$ corresponding to all trivial clusters. Under this provision, there is always at least one  non-trivial cluster whose stability can be studied.}




We can assess CS stability \cite{pecora:2014cluster} by studying small perturbations about the cluster synchronous solution $\bw_i(t)=(\bx_i(t)-\bs_{{i*}}^\sigma(t)))$, where $\bs_{{i*}}^\sigma(t)$ indicates the synchronous solution of the cluster $\mathcal{C}_k$ to which node $i$ belongs \textcolor{black}{(the superscript $\sigma$ is used to emphasize that the cluster synchronous solution depends on $\sigma$ through Eq.\ \eqref{net:Q}}),
\begin{equation} \label{net:perturb}
    \dot{\bw}_i(t)=D\bF(\bs_{{i*}}^\sigma(t)) {\bw}_i(t)+\sigma \sum_j \tilde{A}_{ij} D \bH(\bs_{j*}^\sigma(t)) {\bw}_j(t),
\end{equation}
$i=1,...,N$.


The concept of isolated CS, originally introduced in \cite{pecora:2014cluster}, indicates that the range of a given parameter that synchronizes
each cluster may be different for different clusters.
\begin{definition} \label{def:isolated} \textbf{Isolated Cluster Synchronization.}  
Isolated synchronization occurs when for a given coupling strength, $\sigma$, \textcolor{black}{nodes in} one or more clusters synchronize but \textcolor{black}{nodes in} others \textcolor{black}{clusters} do not.
\end{definition}

\section{Cluster Synchronizability} \label{MSF&Synch}

In what follows we categorize the networks to have either:
\begin{enumerate} [label=(\alph*)]
  \item no intertwined clusters, or
  \item at least one intertwined cluster.
\end{enumerate}
The concept of intertwined clusters was originally introduced in \cite{pecora:2014cluster}. In Sec.\ II of the Supplementary Information we provide a rigorous definition of intertwined clusters.
In this paper, we mostly focus on case (a), which we present in section \ref{sec:boundedNOintwnd}. We briefly present case (b) in Sec.\,\ref{intertwined}.

\subsection{Networks with no intertwined clusters} \label{sec:boundedNOintwnd}


Now we discuss the general case of cluster synchronizability in networks with no intertwined clusters. 
 We \textcolor{black}{omit the derivation} and present the main result that by using a proper similarity transformation $T$\cite{pecora:2014cluster,panahi2021cluster}, the coupling matrix $\tilde{A}$ can be decomposed as follows,
\begin{equation} \label{Tmatrix}
    T \tilde{A} T^{-1}=Q \bigoplus R,
\end{equation}
where the symbol $\bigoplus$ indicates the direct sum of matrices, the `quotient' matrix $Q$ is $K$-dimensional and the `transverse' matrix $R$ is $(N-K)$-dimensional diagonal matrix. \textcolor{black}{A proof of Eq.\ \eqref{Tmatrix} can be found in Sec.\ IV of the Supplementary Material.} The entries on the main diagonal of $R$ are the `transverse eigenvalues' of $\tilde{A}$. One can immediately see that the set $\Lambda$ of the eigenvalues of the matrix $\tilde{A}$ is partitioned into the two sets: the set $\Lambda_Q = \{\lambda_1^Q,...,\lambda_K^Q\}$ of the quotient eigenvalues of the matrix $Q$ and the set  $\Lambda_R = \{\lambda_1^R,...,\lambda_{(N-K)}^R\}$ of the transverse eigenvalues of the matrix $R$, $\Lambda= \Lambda_Q \cup \Lambda_R$, and $\Lambda_Q \cap \Lambda_R =\emptyset$.

The set of transverse eigenvalues, $\Lambda_R$, can further be partitioned into $K$ subsets, $\Lambda_{R}^k$, $k = 1, ..., K$, where each subset corresponds to one cluster\cite{sorrentino2016approximate,panahi2021cluster} \textcolor{black}{(for more details see Sec.\ IV the Supplementary Information)}. The subset $\Lambda_{R}^k$ contains $(n_k - 1)$ eigenvalues $\lambda_{R,i}^{k}$, where $n_k$ is the number of nodes in cluster $k$, and $i = 1, ..., n_k - 1$. Since here we are only concerned with the transverse eigenvalues, we omit the subscript $R$ and indicate the $i$th transverse eigenvalue from cluster $k$ with the notation $\lambda_i^k$, i.e., $\lambda_i^k \in \Lambda_{R}^k$, $i = 1, ..., n_k - 1$, and $k = 1, ..., K$. 
We also label $\mu_i^k=-\lambda_i^k \geq 0$. Then, $\mu_{\max}^k=\max_i \mu_i^k$ and $\mu_{\min}^k=\min_i \mu_i^k$ for each cluster $\mathcal{C}_k$, $k = 1, ..., K$. We will use this notation to define the cluster synchronizability later on.

As a result, the set of equations \eqref{net:perturb} can be transformed into a set of  independently evolving perturbations, which can be divided into two categories, those parallel to the synchronization manifold $\by^p(t) \in \mathbb{R}^{Km}$ and those transverse $\by^t(t)$ to the synchronization manifold.
Only the transverse perturbations $\by^t(t) \in \mathbb{R}^{(N-K)m}$ determine stability about the cluster synchronous solution. Following \cite{sorrentino2016approximate}, the dynamics of the transverse perturbations can be written as,
\begin{equation} \label{pertt}
    \dot{\by}^t_i(t)=\left[D\bF({\bs}_{i*}^\sigma(t)) -\sigma \mu_{i}^R D\bH(\bs_{i*}^\sigma(t))\right] {\by}^t_i(t),
\end{equation}
$i=1,...,(N-K)$, where in Eq.\ \eqref{pertt} for convenience we use \emph{the negative of the eigenvalues} $\mu_i^R=-\lambda_i^R$. Based on our assumption A1, which we introduce below, all the $\lambda_i^R \leq 0$, and so all the $\mu_i^R \geq 0$. The discussion that follows could use either the eigenvalues $\lambda_i^R$ or the negative of the eigenvalues $\mu_i^R$. We opt for the $\mu_i^R$ as in general we find easier to refer to positive quantities.


\color{black}

\begin{definition}\textbf{Stability of the Synchronous State of each Cluster.}
The synchronous state of each cluster $i*=1,...,K$ is stable if the maximum Lyapunov exponents associated with Eq.\ \eqref{pertt} is negative for all $i \in \Lambda_R^{i*}$, i.e., if $MLE({s_{i*}^\sigma},\sigma \mu_{i}^{R})<0$, for all $i \in \Lambda_R^{i*}$,  where the first argument is used to indicate the dependence on the cluster synchronous solution $\bs_{i*}^\sigma(t)$ (which depends on $\sigma$) and the second argument is used to indicate the dependence on the products $ \mu_{i}^{R}$ times $\sigma$.
\end{definition}
\begin{remark}
The fact that $MLE({s_{i*}^\sigma}, \sigma \mu_{i}^{R})<0$ depends on $\sigma$ through both its arguments is in contrast to the case of complete synchronization for which the synchronous solution does not depend on $\sigma$ \cite{Pe:Ca}.
\end{remark}

\begin{definition}\textbf{Cluster Master Stability Function.}
We can now introduce the parameter $\xi= \mu_{i}^{R}$ and define the cluster master stability function (CMSF) $\mathcal{M}_{k}^{\sigma}(\sigma \xi)$ which returns the MLE associated with Eq.\ \eqref{pertt} corresponding to the cluster synchronous solution $\bs_{k}^\sigma(t)$ as a function of the parameter $\sigma \xi$, $i \in \Lambda^k_R$.
\end{definition}

\begin{remark}
Note that the cluster master stability function $\mathcal{M}_{k}^{\sigma}(\sigma \xi)$ depends on $\sigma$  through both the cluster synchronous solution $\bs_{k}^\sigma(t)$ (superscript $\sigma$) and the argument $\sigma \xi$.
\end{remark}


In what follows we proceed under the following three assumptions:
\begin{enumerate} \label{assump}

\itema{The eigenvalues of the matrix $R$ are all negative, which can be ensured by appropriately choosing $\kappa$.} 

\itemb{\textcolor{black}{The CMSF is either negative in a bounded interval of its argument,  i.e.,  $\mathcal{M}^{\sigma}_{k}(\sigma \xi)<0$ for $a_1^{k,\sigma}\leq \xi \leq a_2^{k,\sigma}$, where $a_1^{k,\sigma}, a_2^{k,\sigma} > 0$ (equivalently, $b_1^{k,\sigma}\leq \sigma \xi \leq b_2^{k,\sigma}$, with $b_1^{k,\sigma}=\sigma a_1^{k,\sigma}$ and  $b_2^{k,\sigma}=\sigma a_2^{k,\sigma}$); or in an unbounded range of its argument, i.e.,  $\mathcal{M}^{\sigma}_{k}(\sigma \xi)<0$ for $a_c^{k,\sigma}\leq \xi$, where $a_c^{k,\sigma} > 0$ (equivalently, $b_c^{k,\sigma}\leq \sigma \xi$, with $b_c^{k,\sigma}=\sigma a_c^{k,\sigma}$.) In what follows, we will simply refer to the former as `the bounded case' and to the latter as `the unbounded case'.}}

\itemc{\textcolor{black}{The lower and upper bounds $a_1^{k,\sigma}$ and $a_2^{k,\sigma}$ vary smoothly with $\sigma$.}}
\end{enumerate}


\color{black}

{
Following the discussion in Sec.\ \ref{CompSynch}, we can then introduce the following definitions for the synchronizability of each cluster:
\begin{definition}\textbf{Synchronizability of each cluster.}
In the bounded case, the synchronizability of cluster $k$ is equal to the eigenratio $\eta_b^{k}=\dfrac{\mu_{min}^{k}}{\mu_{max}^{k}}$, and in the unbounded case it is equal to $\eta_u^{k}={\mu_{min}^{k}}$, where $\mu_{min}^{k}$ ($\mu_{max}^{k}$) is the smallest (largest) eigenvalue in $\Lambda_R^{k}$.
\end{definition}}

\begin{figure}[H]
    \centering
    \includegraphics[scale=.4]{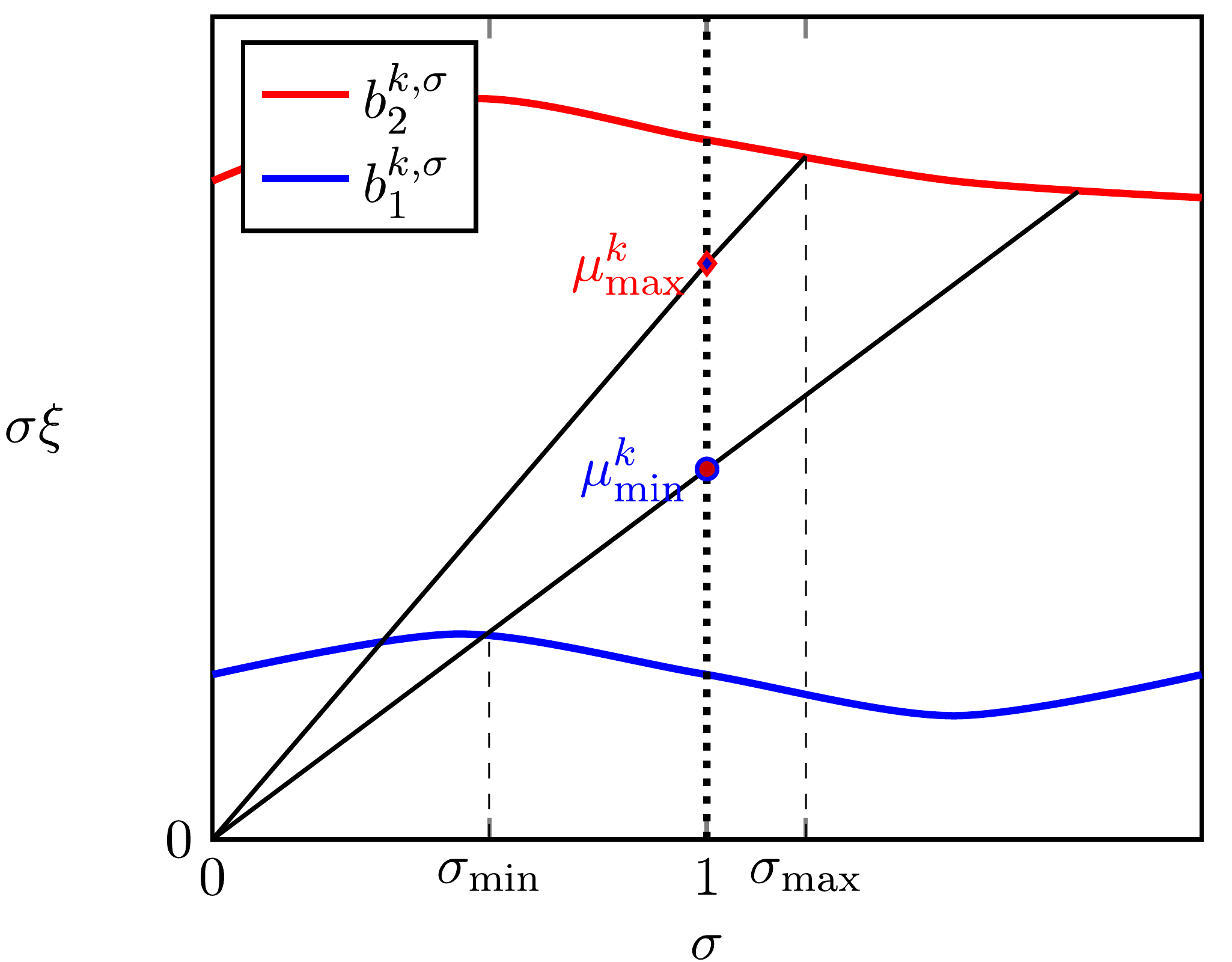}
    \caption{Schematic representation of the synchronizability of cluster $\mathcal{C}_k$ for the bounded case. The master stability function $\mathcal{M}^{\sigma}_{k}(\sigma \xi)<0$ for $b_1^{k,\sigma}\leq \sigma\xi \leq b_2^{k,\sigma}$. The figure shows the lines 
    $\sigma \mu_{\max}^k$ and $\sigma \mu_{\min}^k$, going through the maximum and minimum eigenvalue, $\xi = \mu_{\max}^k$ and $\xi = \mu_{\min}^k$ at $\sigma=1$ (there can be other eigenvalues in between.)  The lines intersect the curves $b_1^{k,\sigma}$ and  $b_2^{k,\sigma}$ twice. However, $\sigma_{\min}$ is determined by the intersection of $\sigma \mu_{\min}^k$ with $b_1^{k,\sigma}$
    and $\sigma_{\max}$ is determined by the intersection of $\sigma \mu_{\max}^k$ with $b_2^{k,\sigma}$.
    The critical coupling strengths, $\sigma_{\min}$ and $\sigma_{\max}$, determine the width of the $\sigma$-interval over which the synchronous solution is stable. }
    \label{fig:synchronizability}
\end{figure}

\begin{remark}
The definitions of synchronizability of cluster $k$ are consistent with the observation that given $b_1^{k,\sigma}$ and $b_2^{k,\sigma}$, increasing $\eta^k_b$ ($\eta^k_u$) will necessarily increase the cluster synchronizability, i.e., the range of $\sigma$ over which the CS solution is stable. This is explained graphically in Fig.\ \ref{fig:synchronizability} for the bounded case.
\end{remark}

 \textcolor{black}{Figure \ref{fig:synchronizability}  shows how the eigenratio $\mu_{\min}^k / \mu_{\max}^k$ relates to the width of the $\sigma$-interval in which  cluster $k$ synchronizes.  The master stability function $\mathcal{M}^{\sigma}_{k}(\sigma \xi)<0$ for $b_1^{k,\sigma}\leq \sigma\xi \leq b_2^{k,\sigma}$, where $b_1^{k,\sigma}$ and $b_2^{k,\sigma}$ are two arbitrary functions of $\sigma$. The figure shows the lines 
    $\sigma \mu_{\max}^k$ and $\sigma \mu_{\min}^k$, 
    intersecting the functions $b_1^{k,\sigma}$ and  $b_2^{k,\sigma}$ twice. However, independent of the particular shapes of the functions $b_1^{k,\sigma}$ and  $b_2^{k,\sigma}$, $\sigma_{\min}$ is determined by the intersection of the line $\sigma \mu_{\min}^k$ with $b_1^{k,\sigma}$
    and $\sigma_{\max}$ is determined by the intersection of the line $\sigma \mu_{\max}^k$ with $b_2^{k,\sigma}$.
    The critical coupling strengths, $\sigma_{\min}$ and $\sigma_{\max}$, determine the range of $\sigma$ over which the synchronous solution of cluster $k$ is stable. We can then write $\dfrac{\sigma_{\max}}{\sigma_{\min}} = \dfrac{b_{2}^{k,\sigma}}{b_{1}^{k,\sigma}} \dfrac{\mu_{\min}^k}{\mu_{\max}^k}$, where the first ratio $\dfrac{\mu_{\min}^k}{\mu_{\max}^k}$ represents the effect of the network topology and the second ratio $ \dfrac{b_{2}^{k,\sigma}}{b_{1}^{k,\sigma}}$ represents the effect of the dynamics, in terms of the functions $\bF$ and $\bH$. As the ratio $\dfrac{\mu_{\min}^k}{\mu_{\max}^k}$ increases, i.e., as the eigenvalues move closer to one another, the synchronizability of cluster $k$ increases. Changing the functions $\bF$ and $\bH$ only moves the functions ${b_{2}^{k,\sigma}}$ and ${b_{1}^{k,\sigma}}$ (and not the lines $\sigma \mu_i^k$). Changing the network topology affects the lines $\sigma \mu_i^k$, but also, through the dynamics of the quotient network, the functions $ {b_{2}^{k,\sigma}}$ and ${b_{1}^{k,\sigma}}$. Our main result is that even if these two functions change as a result of changing the network topology, the intersections with the lines $\sigma \mu_i^k$ always occur for $\mu_i^k=\mu_{\min}^k$ and  $\mu_i^k=\mu_{\max}^k$. }

\color{black}
\color{black}

\color{black}

\textcolor{black}{So far we have proposed a definition of cluster synchronizability that independently applies to each cluster within a complex network. Next we study conditions for the entire network to synchronize, i.e., for nodes in several clusters to synchronize at the same time. 
Since nodes in one cluster may synchronize independently of nodes in another cluster, this type of analysis is considerably more complex and will lead to the observation of different \emph{types} of cluster synchronization, namely, Matryoshka CS, Partially Disjoint CS and Completely Disjoint CS. Under general conditions,  we will not be able to derive a definition of cluster synchronizability for the entire network that reflects the network structure. Both the bounded case and the unbounded case require considerations that generally depend on both the network structure and the dynamics.} 

\textcolor{black}{The bounded case is simpler, thus it is discussed first. In this case, cluster $k$ synchronizes for $\sigma>b_c^{k,\sigma_{\min}}/ \mu_{\min}^k $. Therefore, all nontrivial clusters synchronize for $\sigma>\max_{k=1,...,\tilde{K}} \frac{b_c^{k,\sigma_{\min}}}{ \mu_{\min}^k} $.
\begin{definition} \textbf{Network Cluster Synchronizability (unbounded case.)} \label{clustsynchnetunbd}
The cluster synchronizability of a network $\eta_u$ is equal to
\begin{equation}  \label{etau:General}
\eta_u \coloneqq
\max_{k=1,...,\tilde{K}} \frac{b_c^{k,\sigma_{\min}}}{ \mu_{\min}^k}.
\end{equation}
The larger $\eta_u$, the more cluster synchronizable a network.
\end{definition}
As can be seen, this definition does not decouple the dynamics from the network structure.
The rest of this section is devoted to the bounded cases, which is more complex.}

Generally \textcolor{black}{in the bounded case}, for each cluster $k$ an interval of $\sigma$, $[\sigma_{\min}^k, \sigma_{\max}^k]$  can be found such that the synchronous solution of that particular cluster is stable. The intervals for stability of the synchronous solution of each cluster are not necessarily the same, which is associated with the phenomenon of Isolated CS \cite{pecora:2014cluster}. 

\begin{definition} \textbf{Condition for Cluster Synchronization of the Network (bounded case.)} \label{CsynchnetDEF}
A network can be cluster-synchronized if there exists a value of $\sigma$ such that 
\begin{equation} \label{CsynchNet}
  \max_{k = 1, ..., \tilde K} \left(\sigma_{\min}^k \right) < \sigma < \min_{k = 1, ..., \tilde K} \left( \sigma_{\max}^k \right).
\end{equation}
\end{definition}
Note that the definition \ref{CsynchnetDEF} corresponds to all nontrivial cluster synchronizing.  
Now, let
\begin{equation} \label{GenEqCSynchclust}
    k_1 =  \textcolor{black}{\operatorname*{arg\,max}_{k = 1, ..., \tilde{K}}}\left (\sigma_{\min}^k \right), \quad k_2 = \textcolor{black}{\operatorname*{arg\,min}_{k = 1, ..., \tilde{K}}} \left (\sigma_{\max}^k \right),
\end{equation}
\textcolor{black}{where $k_1$ and $k_2$ indicate the critical cluster $\mathcal{C}_{k_1}$ and $\mathcal{C}_{k_2}$ that correspond to the largest lower bound and the smallest upper bound among all the clusters, respectively.} \textcolor{black}{We further clarify what we mean by critical clusters. Let us assume that for a given $\sigma$ value, nodes in all clusters are synchronized in their respective cluster synchronous solutions, $\bs_{k}$, $k = 1, ..., K$. In the case of bounded MSF, as we decrease (increase) this $\sigma$, there will be one critical cluster $\mathcal{C}_{k_1}$ ($\mathcal{C}_{k_2}$) that its nodes are the first to desynchronize. }
In general, $k_1$ and $k_2$ can be either the same ($k_1 = k_2$) or different ($k_1 \neq k_2$). In either case, we can rewrite Eq. \eqref{CsynchNet} as,
\begin{equation}
    \sigma_{\min}^{k_1}\leq \sigma \leq \sigma_{\max}^{k_2}.
\end{equation}
\begin{definition} \textbf{The Network Cluster Synchronizability Ratio (bounded case.)} \label{eta:MostGeneral}
The network cluster synchronizability ratio $\rho$ is defined as the ratio between the maximum and minimum values of $\sigma$ for which the network cluster-synchronizes,
\begin{equation} \label{eta}
    \rho \coloneqq \dfrac{\sigma_{\max}^{k_2}}{\sigma_{\min}^{k_1}}.
\end{equation}
\end{definition}

{The ratio $\rho$ can be directly measured in simulation by integrating Eq.\ \eqref{net:Aclust} for different values of the parameter $\sigma$ and recording the smallest and the largest values of $\sigma$ for which CS is achieved.} The cluster synchronizability ratio, $\rho$, measures how large the $\sigma$-interval is over which the network cluster-synchronizes.
Under the assumption that clusters are not intertwined, Eq.\ \eqref{eta} can be rewritten as,

\textcolor{black}{
\begin{equation}
    \rho = \dfrac{\sigma_{\max}^{k_2}}{\sigma_{\min}^{k_1}} ={\dfrac{b_{2}^{k_2, \sigma}}{b_{1}^{k_1, \sigma}}} {\dfrac{{\mu_{\min}^{k_1}}}{\mu_{\max}^{k_2}}},
\end{equation}}
where the ratio ${b_{2}^{k_2, \sigma}}/{b_{1}^{k_1, \sigma}}$ is related to the particular choice of the dynamical functions $\bF$ and $\bH$, and the ratio ${{\mu_{\min}^{k_1}}}/{\mu_{\max}^{k_2}}$ is related to the network topology (with an important caveat which we explain below.) We can then introduce the following definition,
\begin{definition} \textbf{Network Cluster Synchronizability (bounded case.)} \label{clustsynchnet}
The cluster synchronizability of a network $\eta_b$ is equal to the ratio between the minimum and the maximum of the negative of the eigenvalues of the critical clusters $\mathcal{C}_{k_1}$ and $\mathcal{C}_{k_2}$ that result from Eq. \eqref{GenEqCSynchclust}, i.e.,
\begin{equation}  \label{eta:General}
\eta_b \coloneqq
\frac{\mu_{\min}^{k_1}}{\mu_{\max}^{k_2}}.
\end{equation}
The larger $\eta_b$, the more cluster synchronizable the network. 
\end{definition}

\textcolor{black}{We stress that, different from the case of complete synchronization (Defs.\ \ref{ComNetBounded} and \ref{ComNetUnbounded}), the definition of network cluster synchronizability (Eq.\ \eqref{eta:General}) does not solely depend on the network topology. In fact, since the cluster indices $k_1$ and $k_2$ are obtained from Eq. \eqref{GenEqCSynchclust}, we see that the cluster synchronizability of the network, $\eta_b$, also depends on the CMSF bounds, i.e., $b_1^{k, \sigma}$ and $b_2^{k, \sigma}$, $k = 1, ..., \tilde K$. This represents a main difference with the case of complete synchronization \cite{barahona:2002}.} \textcolor{black}{We emphasize that unlike the complete synchronization case, our definition of synchronizability, Eq. \eqref{eta:General} depends on both the topology and the dynamics.} 

 In what follows, we look in more detail at the possible types of cluster-to-cluster synchronization that can be observed in a network and sub-categorize isolated synchronization in the following three types:

\begin{enumerate} [(a)]
  \item Matryoshka Cluster Synchronization, which occurs when the synchronous $\sigma$-interval of cluster $\mathcal{C}_a$ is completely contained in the $\sigma$-interval of cluster $\mathcal{C}_b$, 
\begin{equation}
    \left[\sigma_{\min}^{a},\, \sigma_{\max}^{a}\right] \subseteq  \left[\sigma_{\min}^{b},\, \sigma_{\max}^{b}\right].
\end{equation}
If nodes in cluster $\mathcal{C}_{a}$ synchronizes, nodes in cluster $\mathcal{C}_b$ synchronizes as well, but the reverse implication is not true. 
  \item Partially Disjoint Cluster Synchronization, which occurs when the intersection of the synchronous $\sigma$-intervals of clusters $\mathcal{C}_{a}$ and $\mathcal{C}_b$ is nonempty and it does not coincide with any of the two intervals. In this case, the lower and upper bounds of synchronous coupling strength, $\sigma$, are not from the same cluster:
\begin{equation}\label{ParcriticalClus}
    a =  \textcolor{black}{\operatorname*{arg\,max}_{k = a,b}}\left (\sigma_{\min}^k \right), \quad b = \textcolor{black}{\operatorname*{arg\,min}_{k = a,b}} \left (\sigma_{\max}^k \right).
\end{equation}
  \item Complete  Disjoint Cluster Synchronization, which occurs when the intersection of the synchronous $\sigma$-intervals of two clusters is empty. \label{Com} In this case, the lower bound of $\sigma$ from Eq.\ \eqref{CsynchNet} is greater than the upper bound. As a result, there is no $\sigma$ such that the synchronous solutions of the two clusters are stable at the same time, and we can define $\eta_b\coloneqq0$.
\end{enumerate}
These definitions can be generalized to the case of more than two clusters. 

Given the above definitions, the network cluster synchronizability, $\eta_b$, provides specific information about the type of CS:
\begin{itemize}
    \item A necessary condition for Matryoshka CS is that $0<\eta_b \leq1$.
    \item A sufficient condition for Partially Disjoint CS is that $\eta_b > 1$.
    \item A necessary and sufficient condition for Complete Disjoint CS is that $\eta_b = 0$.
\end{itemize}

\subsubsection{Numerical Analysis}

\begin{figure}[h] 
    \centering
    \includegraphics[scale=.8]{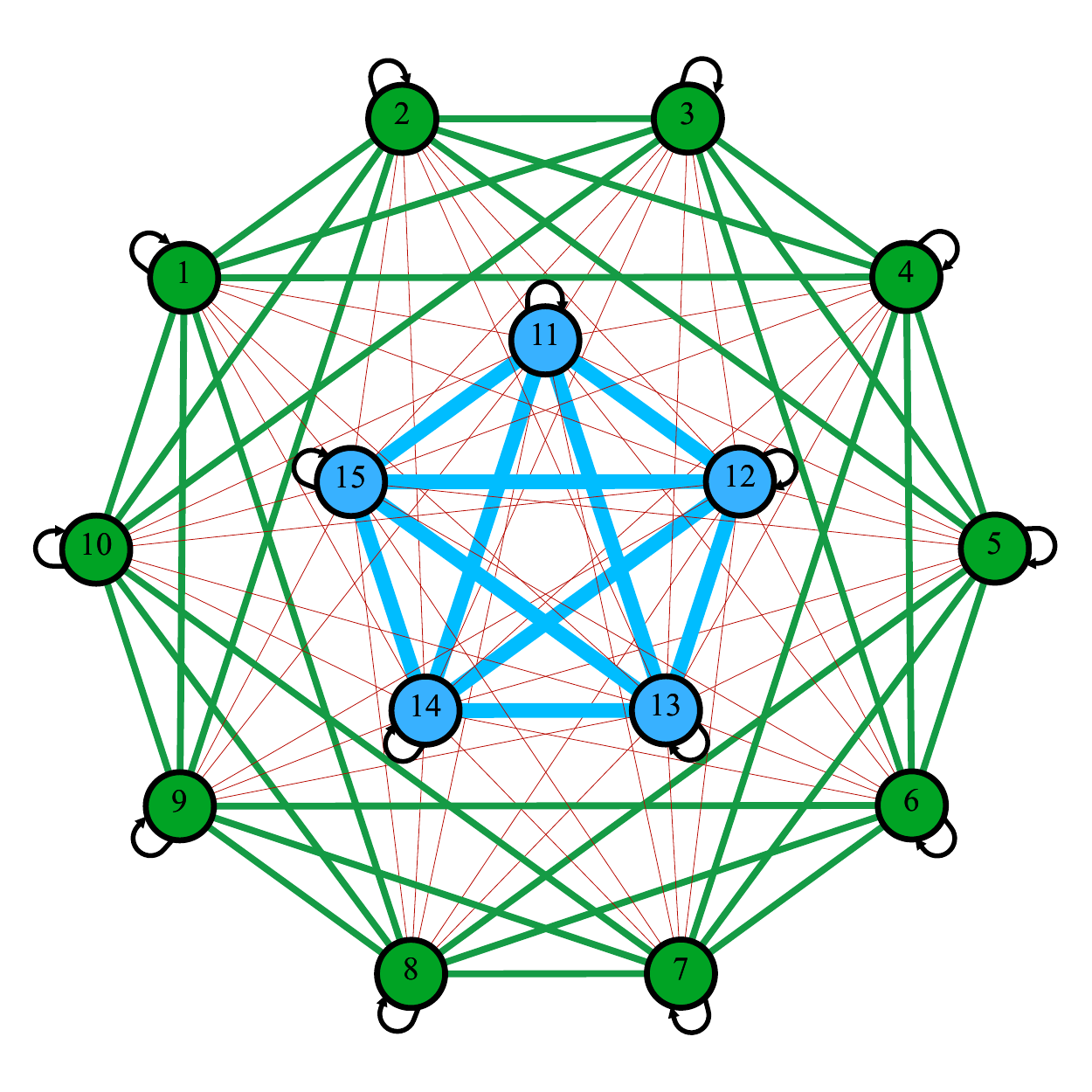}
    \caption{{15-node undirected and weighted network.} The adjacency matrix of this network, $\tilde{A}$, is shown in Eq.\ \eqref{ex1}. Clusters are shown in different colors: Green nodes (1 through 10) are in the first cluster, $\mathcal{C}_a$, and black nodes (11 trough 15) are in the second cluster, $\mathcal{C}_b$. Green, black, red, and black edges have weights $w_a$, $w_b$, $w_c$, and, $-\kappa$, respectively.}
    \label{fig:graph15}
\end{figure}

As a reference example, we consider the network with $N=15$ nodes and $K=2$ clusters in Fig.\ \ref{fig:graph15}.  This network consists of two \textit{circles} corresponding to two clusters: the larger cluster, $\mathcal{C}_a$, has $n_a = 10$ nodes where each node is connected to its six nearest neighbors, with coupling weight $w_a$. The smaller cluster, $\mathcal{C}_b$, is a fully connected {circle} with $n_b = 5$ nodes and coupling weight $w_b$. The two circles are fully connected with one another with coupling weight $w_c$. The shifted adjacency matrix, $\tilde{A} = A - \kappa I_{15}$, is,
\begin{equation} 
\resizebox{0.65\hsize}{!}{$\tilde{A}=\left [
\begin{array}{*{16}{@{}C{\mycolwd}@{}}}
-\kappa & w_a & w_a & w_a & 0 & 0 & 0 & w_a & w_a & w_a & \vline & w_c & w_c & w_c & w_c & w_c \\ 
w_a & -\kappa & w_a & w_a & w_a & 0 & 0 & 0 & w_a & w_a & \vline & w_c & w_c & w_c & w_c & w_c \\ 
w_a & w_a & -\kappa & w_a & w_a & w_a & 0 & 0 & 0 & w_a & \vline & w_c & w_c & w_c & w_c & w_c \\ 
w_a & w_a & w_a & -\kappa & w_a & w_a & w_a & 0 & 0 & 0 & \vline & w_c & w_c & w_c & w_c & w_c \\ 
0 & w_a & w_a & w_a & -\kappa & w_a & w_a & w_a & 0 & 0 & \vline & w_c & w_c & w_c & w_c & w_c \\ 
0 & 0 & w_a & w_a & w_a & -\kappa & w_a & w_a & w_a & 0 & \vline & w_c & w_c & w_c & w_c & w_c \\ 
0 & 0 & 0 & w_a & w_a & w_a & -\kappa & w_a & w_a & w_a & \vline & w_c & w_c & w_c & w_c & w_c \\ 
w_a & 0 & 0 & 0 & w_a & w_a & w_a & -\kappa & w_a & w_a & \vline & w_c & w_c & w_c & w_c & w_c \\ 
w_a & w_a & 0 & 0 & 0 & w_a & w_a & w_a & -\kappa & w_a & \vline & w_c & w_c & w_c & w_c & w_c \\ 
w_a & w_a & w_a & 0 & 0 & 0 & w_a & w_a & w_a & -\kappa & \vline & w_c & w_c & w_c & w_c & w_c \\ 
\hline
w_c & w_c & w_c & w_c & w_c & w_c & w_c & w_c & w_c & w_c & \vline & -\kappa & w_b & w_b & w_b & w_b \\ 
w_c & w_c & w_c & w_c & w_c & w_c & w_c & w_c & w_c & w_c & \vline & w_b & -\kappa & w_b & w_b & w_b \\ 
w_c & w_c & w_c & w_c & w_c & w_c & w_c & w_c & w_c & w_c & \vline & w_b & w_b & -\kappa & w_b & w_b \\ 
w_c & w_c & w_c & w_c & w_c & w_c & w_c & w_c & w_c & w_c & \vline & w_b & w_b & w_b & -\kappa & w_b \\ 
w_c & w_c & w_c & w_c & w_c & w_c & w_c & w_c & w_c & w_c & \vline & w_b & w_b & w_b & w_b & -\kappa
\end{array}
\right ].$} \label{ex1}
\end{equation}
{In order to assess the emergence of cluster synchronization, Eq.\ \eqref{net:Aclust} is integrated for a wide range of the coupling strength $\sigma$. For each value of $\sigma$, the initial conditions were determined as follows. First, the dynamics of the uncoupled system ${\bs(t)}$ is integrated for a long time so that it converges on its attractor. Then, a randomly selected point from the attractor of ${\bs(t)}$ is picked and set as the initial condition of the quotient equation, Eq. \eqref{net:Q}. For a given value of $\sigma$, the quotient network dynamics \eqref{net:Q} is integrated for a long time so that it converges on its attractor.
Then, a point on the synchronous state for each cluster, ${\bs_k(t)}$, $k = 1, ..., K$, is picked randomly, and the following steps are taken:
 \begin{enumerate}
     \item Small perturbations are added to this point to be used as initial conditions for the network dynamics, Eq. \eqref{net:Aclust} with the selected $\sigma$,
     \item This point is used as the initial condition of the quotient equation, Eq. \eqref{net:Q}, for the next value of $\sigma$.
 \end{enumerate}
The procedure is repeated for increasing values of $\sigma$.}


 We select the individual uncoupled systems to be Van der Pol oscillators,
  \begin{equation} \label{FH}
         \bF(\bx) = \begin{bmatrix}x_{2} \\ 
     -x_{1} + 3(1-x_{1}^2)x_{2}\end{bmatrix}, \quad      \bH(\bx) = \begin{bmatrix} 0\\x_2 \end{bmatrix},
  \end{equation}
 with $\bx = \left[x_1 \quad x_2\right]^T \in \mathbb{R}^2$, $\bF : \mathbb{R}^2 \rightarrow \mathbb{R}^2$, $\bH : \mathbb{R}^2 \rightarrow \mathbb{R}^2$. {Our choice of the functions $\bF$ and $\bH$ is such that the MSF is bounded.} {To find the synchronous coupling strength $\sigma$, we simulate Eq.\ \eqref{net:Aclust} over a wide range of $\sigma$. Then, the synchronization error for each cluster, $E_k$, $k = a,b$, is calculated as,
 \begin{equation}
     E_k = \texttt{<}{\sum_{i = 1}^{n_k} \|\bx_i(t) - \bx_1(t) \|_2}\texttt{>}_t, \quad k = a,b,
 \end{equation}
where $\texttt{<} \cdot \texttt{>}_t$ indicates an average over the time interval $[450, 500]$. The synchronization error represents the time-average of the sum of the norm of the error of each node's states with respect to one of the nodes in the cluster (here we choose node 1.)}

\begin{figure}[h!]
     \centering
     \subfigure[]{\includegraphics[width=0.49\linewidth]{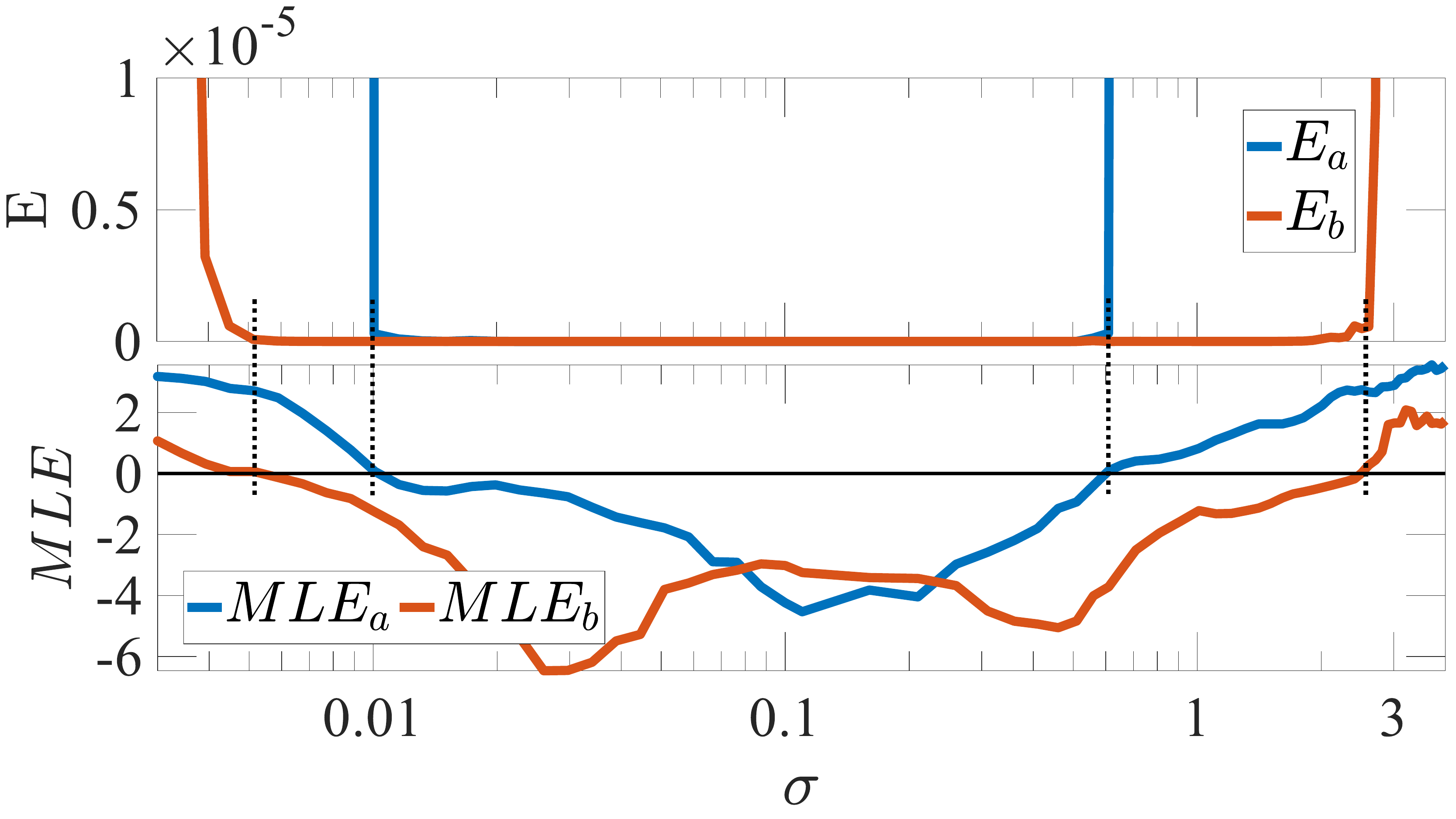}}
     \subfigure[]{\includegraphics[width=0.49\linewidth]{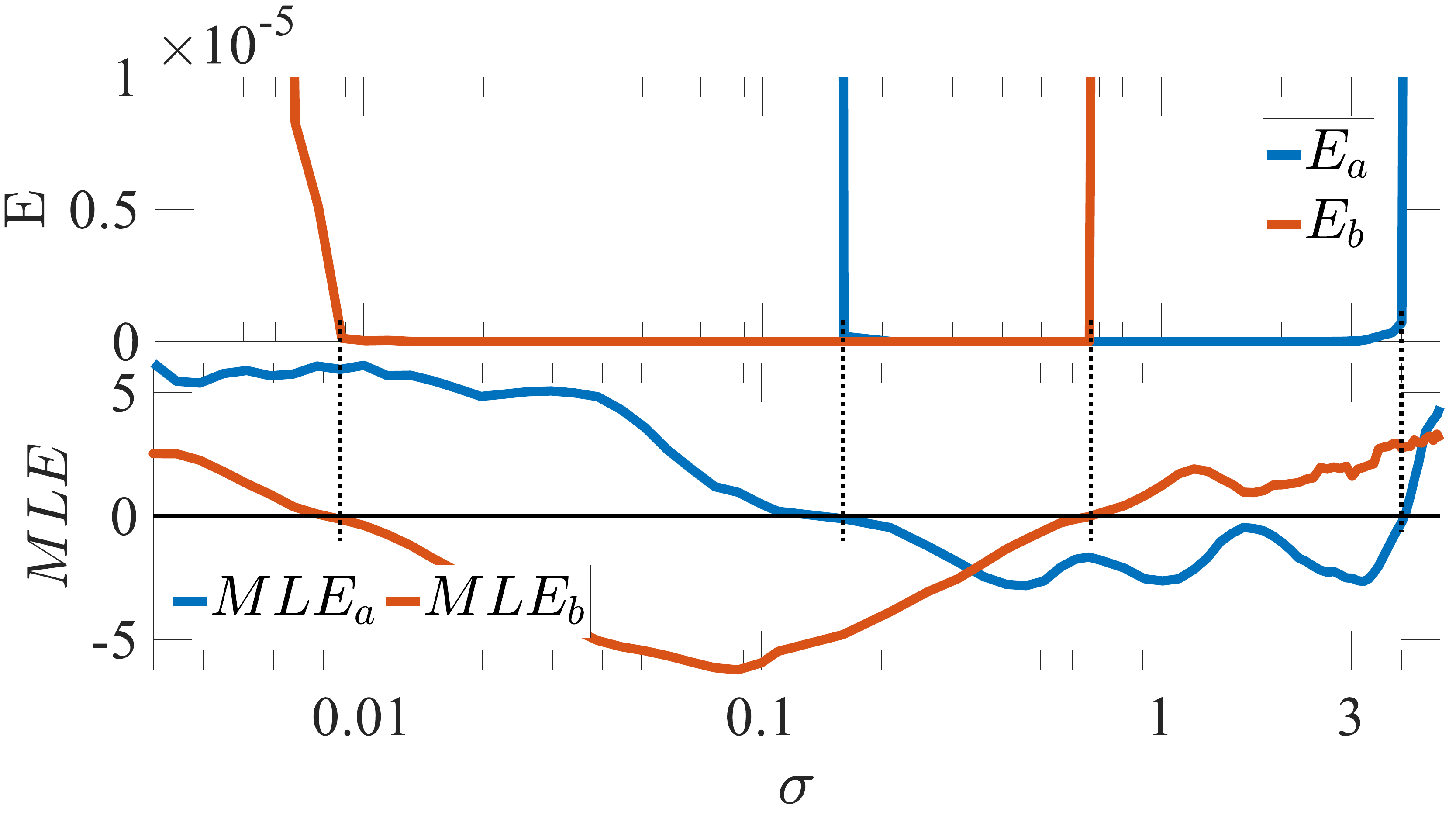}}
     \subfigure[]{\includegraphics[width=0.49\linewidth]{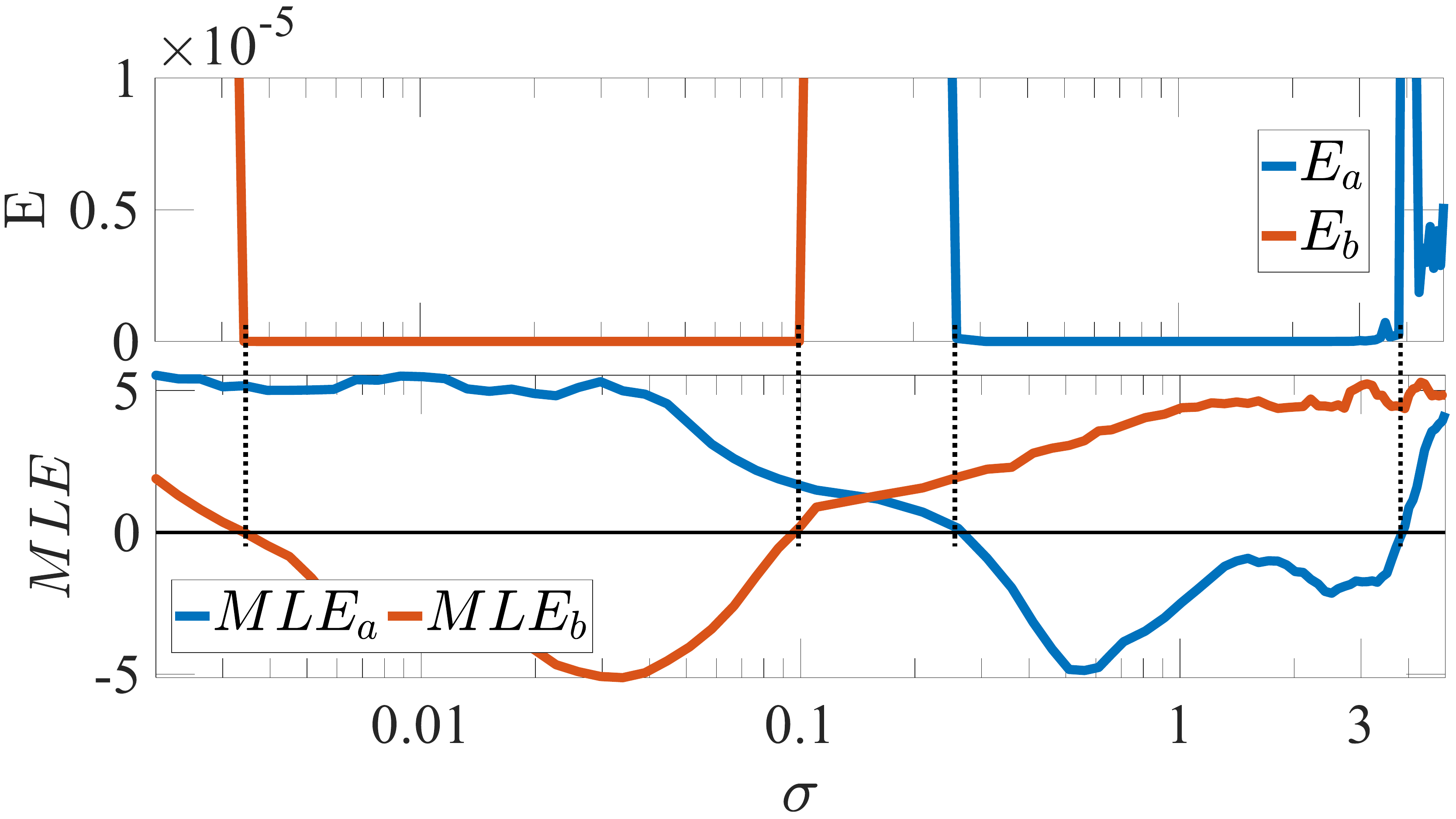}}
\caption{{Isolated CS: Matryoshka, Partially, and Complete Disjoint CS.} In each panel, we plot the CS error $E$,  \textcolor{black}{as well as the maximum Lyapunov exponent $MLE_a$ ($MLE_b$) computed as $\max_{i \in \Lambda^a_R} \mathcal{M}_k^\sigma (\sigma \mu_i^R)$ ($\max_{i \in \Lambda^b_R} \mathcal{M}_k^\sigma (\sigma \mu_i^R)$), as a function of the coupling strength $\sigma$.} 
black and red curves represent the synchronization error, $E_a$ and $E_b$, and also  for the clusters $\mathcal{C}_a$ and $\mathcal{C}_b$. (a) shows the synchronous $\sigma$ interval of cluster $\mathcal{C}_a$ which is contained in the synchronous $\sigma$ interval of cluster $\mathcal{C}_b$, which is Matryoshka synchronization. The synchronous $\sigma$ interval for $\mathcal{C}_a$ is, $0.010 < \sigma <  0.610$, and for $\mathcal{C}_b$ is, $0.005 < \sigma < 2.610$. (b) shows a synchronous $\sigma$ interval in the case that the lower bound is from $\mathcal{C}_a$, and the upper bound is from $\mathcal{C}_b$, which is Partially Disjoint synchronization. The synchronous $\sigma$ interval for $\mathcal{C}_a$ is, $0.160 < \sigma < 4.010$, and for $\mathcal{C}_b$ is, $0.009 < \sigma < 0.66$. (c) shows that there is no synchronous $\sigma$ that synchronizes both clusters at the same time, which is Complete Disjoint synchronization. The synchronous $\sigma$ interval for $\mathcal{C}_a$ is, $0.260 < \sigma < 3.810$, and for $\mathcal{C}_b$ is, $0.003 < \sigma < 0.100$.}
     \label{fig:example}
 \end{figure}

(a) By setting the network parameters to $w_a = 2$, $w_b = 3$, $w_c = 0.1$, $\kappa = 12$, we obtain a case of Matryoshka synchronization. The transverse eigenvalues for the two clusters are $\Lambda_R^a = $ \{-17.24, -17.24, -16, -13.24, -13.24, -12.76, -12.76, -8.76, -8.76\} and $\Lambda_R^b = $ \{-15, -15, -15, -15\}. \textcolor{black}{The upper part of Fig. \ref{fig:example}(a) is a plot of the synchronization error $E_a$ for cluster $\mathcal{C}_a$ (in black) and $E_b$ for cluster $\mathcal{C}_b$ (in red.) The lower part of the figure shows the largest transverse MLE for $\mathcal{C}_a$ (in black) and for $\mathcal{C}_b$ (in red) as a function of $\sigma$ for the transverse motions equation, Eq.\ \eqref{pertt}.} These are computed as $\max_{i \in \Lambda^a_R} \mathcal{M}_k^\sigma (\sigma \mu_i^R)$ and $\max_{i \in \Lambda^b_R} \mathcal{M}_k^\sigma (\sigma \mu_i^R)$, respectively. The $\sigma$-interval for the network synchronization is $0.010 < \sigma <  0.610$. The lower bound and the upper bound are determined by $\mathcal{C}_a$. Therefore, the synchronizability based on Eq. \eqref{eta:General} is
$\eta_b = {\mu_{\min}^{a}}/{\mu_{\max}^{a}} = {8.76}/{17.24} \simeq 0.50.$



(b) Now consider the same network with parameters, $w_a = 0.1$, $w_b = 3$, $w_c = 0.1$, $\kappa = 12$. Then, the transverse eigenvalues for each cluster are $\Lambda_R^a = $ \{-12.26, -12.26, -12.20, -12.06, -12.06, -12.04, -12.04,  -11.84, -11.84\} and $\Lambda_R^b = $\{-15, -15, -15, -15\}. This corresponds to a case of Partially Disjoint synchronization (see Fig.\ \ref{fig:example}(b)). \textcolor{black}{The upper part of Fig. \ref{fig:example}(b) is a plot of the synchronization error $E_a$ for cluster $\mathcal{C}_a$ (in black) and $E_b$ for cluster $\mathcal{C}_b$ (in red.) The lower part of the figure shows the largest transverse MLE for $\mathcal{C}_a$ (in black) and for $\mathcal{C}_b$ (in red) as a function of $\sigma$ for the transverse motions equation, Eq.\ \eqref{pertt}.}  The  $\sigma$ interval for network synchronization is $0.160 < \sigma <  0.660$. The lower bound and the upper bound are determined from the first and second clusters, respectively. Therefore, the synchronizability based on Eq. \eqref{eta:General} is $\eta_b = {\mu_{\min}^{a}}/{\mu_{\max}^{b}} = {11.84}/{15} \simeq 0.79.$

(c) Finally, we consider the network parameters $w_a = 0.1$, $w_b = 10$, $w_c = 0.1$, $\kappa = 12$. Then, the transverse eigenvalues of each cluster are $\Lambda_R^a = $ \{-12.26, -12.26, -12.20, -12.06, -12.06, -12.04, -12.04,  -11.84, -11.84\} and $\Lambda_R^b = \{-22, -22, -22, -22\}$. This corresponds to a case of Complete Disjoint synchronization, for which the synchronizability is then equal to $\eta_b = 0$. \textcolor{black}{The upper part of Fig.\ \ref{fig:example}(c)) is a plot of the synchronization errors for each cluster as a function of $\sigma$. The lower part of the figure shows the largest transverse MLE  for both clusters as a function of $\sigma$ for the transverse motions equation, Eq.\ \eqref{pertt}.}

\begin{figure}[h] 
    \centering
    \includegraphics[scale=.45]{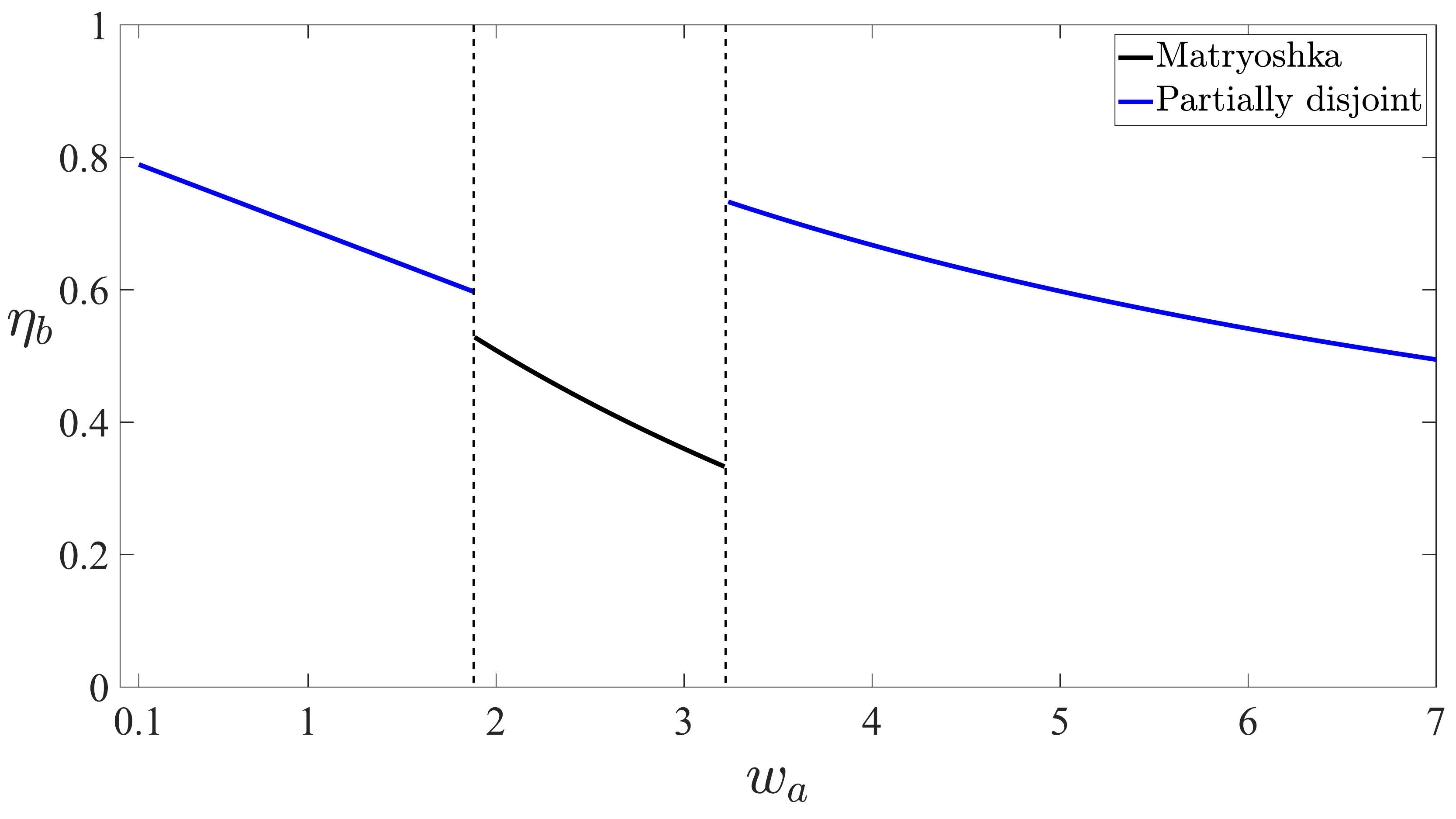}
    \caption{Synchronizability $\eta_b$ as the coupling weight of the first cluster $w_a$ is varied, while the other parameters of the network, $\kappa = 12,\, w_c = 0.1$, $w_b = 3$, are kept constant. We integrate the dynamical Eq.\ \eqref{net:Aclust} with functions from Eq.\ \eqref{FH} and the network from Eq.\ \eqref{ex1}. We see that as $w_a$ is varied, the type of synchronization changes. For $0.1 \leq w_a \leq 1.88$, we see a case of Partially Disjoint synchronization where the lower and upper bounds of $\sigma$ are determined by the first and the second clusters, respectively. For $1.89 \leq w_a \leq 3.22$, we see a case of Matryoshka synchronization where the first cluster is the critical cluster. Finally, for $3.23 \leq w_a \leq 7$, we see a case of Partially Disjoint synchronization again where the lower and upper bounds of $\sigma$ are now determined by the second and the first clusters, respectively.}
    \label{fig:eta}
\end{figure}

Figure \ref{fig:eta} shows the synchronizability $\eta_b$ vs. $w_a$ for a dynamical network described by Eqs.\ \eqref{net:Aclust}, \eqref{FH} and adjacency matrix from Eq.\ \eqref{ex1}. The coupling weight of the first cluster, $w_a$, is varied from 0.1 to 7, while the other network parameters $\kappa = 12,\, w_c = 0.1$, $w_b = 3$ are kept constant. For increasing values of $w_a$ from 0.1, we first see Partially Disjoint synchronization, then Matryoshka, then, Partially Disjoint again. The synchronizability $\eta_b$ is shown to decreases in each one of the intervals.

The reason for the `jumps' in the Fig. \ref{fig:eta} is that for different types of cluster synchronization, $\eta_b$ is computed differently based on Def. \ref{clustsynchnet} since the critical clusters vary with $w_a$. For instance, in the Partially Disjoint case for $0.1 \leq w_a \leq 1.88$, synchronizability is calculated as $\eta_b = {\mu_{\min}^{a}}/{\mu_{\max}^{b}}$. For $1.89 \leq w_a \leq 3.22$ which is Matryoshka synchronization, synchronizability is calculated as $\eta_b = {\mu_{\min}^{a}}/{\mu_{\max}^{a}}$. For $3.23 \leq w_a \leq 7$ which is Partially Disjoint synchronization, synchronizability is calculated as $\eta_b = {\mu_{\min}^{b}}/{\mu_{\max}^{a}}$.

\subsection{Networks with intertwined clusters} \label{intertwined}

In the previous sections, we investigated stability of the synchronous solution, $\bs_k (t)$, $k=1,..,={K}$, for networks with no intertwined clusters. In this section we focus on the general case of networks with intertwined clusters. {We first define the $N$-dimensional diagonal indicator matrices $E_k = \{e_{ij}\}$, $k = 1, ..., K$, such that $e_{ii} = 1$ if node $i$ belongs to the cluster $\mathcal{C}_k$, and $e_{ii} = 0$ otherwise.}
{The stability of the CS solution depends on the dynamics of small perturbations around the synchronous solution, $\bw_i(t)=(\bx_i(t)-\bs_{{i*}}(t)))$. The dynamics of the vector $\bw(t) = \left[{\bw_1(t)}^T, \ {\bw_2(t)}^T, ..., {\bw_N(t)}^T \right]^T$ obeys,
\begin{equation} \label{wgeneral}
    \dot{\bw}(t) = \left[\sum_{k = 1}^K E_k \otimes D\bF(\bs_k(t)) +  \sigma \sum_{k = 1}^K \tilde{A} E_k \otimes D\bH(\bs_k(t))\right]\bw(t),
\end{equation}}
By using a proper transformation matrix, $T$ \cite{pecora:2014cluster,zhang2020symmetry,panahi2021cluster}, the shifted adjacency matrix, $\tilde{A}$, is still decomposed into the form of Eq.\ \eqref{Tmatrix},
where the `quotient' matrix $Q$ is $K$-dimensional, and the `transverse' matrix $R$ is $(N-K)$-dimensional. The matrix $T$ decouples the matrix $\tilde{A}$ into finest blocks \cite{pecora:2014cluster,panahi2021cluster, zhang2020symmetry}, which results in the following block-diagonal structure for the transverse matrix $R$, 
\begin{equation}
    R=\bigoplus_{l = 1}^L R_l,
\end{equation}
where $L$ is the total number of blocks in $R$, $1 \leq L < (N-K)$.

We will analyze the stability of the synchronous solution in terms of the dynamics of the transverse perturbations. {After applying the transformation $T$, we can rewrite Eq.\ \eqref{wgeneral},}
\begin{equation} \label{kron}
    {\dot{\tilde{\by}}(t) = \left[\sum_{k = 1}^K E_k \otimes D\bF(\bs_k(t)) +  \sigma \sum_{k = 1}^K B E_k \otimes D\bH(\bs_k(t))\right]\tilde{\by}(t),}
\end{equation}
{where $B = T\tilde{A}T^{-1} = Q \bigoplus R$. The perturbations can be divided into $\tilde{\by}(t) = \left[{\tilde{\by}_Q(t)}^T, \ {\tilde{\by}_R(t)}^T\right]^T$, where ${\tilde{\by}_R(t)} = \left[{\tilde{\by}_1(t)}^T, \ {\tilde{\by}_2(t)}^T, ..., {\tilde{\by}_L(t)}^T \right]^T$ corresponds to the transverse perturbation equations for each block of $R$. For each block of $R_l$, $l =1, ..., L$ there is an interval of $\sigma$ in which $\mathcal{M}(\sigma^l) <0, \, \forall \sigma^l \in \left[\sigma_{\min}^l, \, \sigma_{\max}^l\right]$.} As a result, the network can synchronize if,
\begin{equation} \label{eq:sigmaIntrtwnd}
    \exists \sigma : \max_{l = 1, ..., L} \left( \sigma_{\min}^l\right) \leq \sigma \leq \min_{l = 1, ..., L }\left( \sigma_{\max}^l\right).
 \end{equation}
Note that this is the most general condition for the synchronization presented in this paper. 


In the case of intertwined clusters, the definition of different cases of synchronization, Matryoshka, Partially Disjoint, and Complete Disjoint from Sec.\,\ref{sec:boundedNOintwnd} are still valid, but with the difference that now those definitions apply to the blocks of the matrix $R$. For each pair of blocks, we can either have Matryoshka synchronization (if the synchronous $\sigma$ intervals are contained in one another), or Partially Disjoint synchronization (if one block determines the lower bound and the other determines the higher bound of the synchronous $\sigma$ interval of the pair), or Completely Disjoint synchronization (if there is no value of $\sigma$ that synchronizes the pair of blocks.)

\subsection{Real Networks Analysis} \label{sec:real}

In this section, we consider several real networks from the literature. 
The first dataset, case16ci, is a  power distribution network\cite{Civanlar1988Distribution, Matpower2020}. The second network, case85,  is a radial power distribution network \cite{DAS1995Simple, Matpower2020}. The third network, backward
circuit, is a neural network, originally from \cite{morone2019symmetry} and was manually repaired in \cite{leifer2021symmetrydriven}. The forth network, macaque-rhesus-brain-2, is a human brain network \cite{bigbrain, nr}. The fifth network, rt-retweet, is a retweet network of  Twitter users based on various social and political hashtags\cite{rossi2012fastclique}.  \textcolor{black}{In Table \ref{table:real}, we provide a summary of the main properties of these five networks. All these networks have intertwined clusters.}

\begin{figure}[h!] 
    \centering
    \subfigure[]{\includegraphics[width=0.49\linewidth]{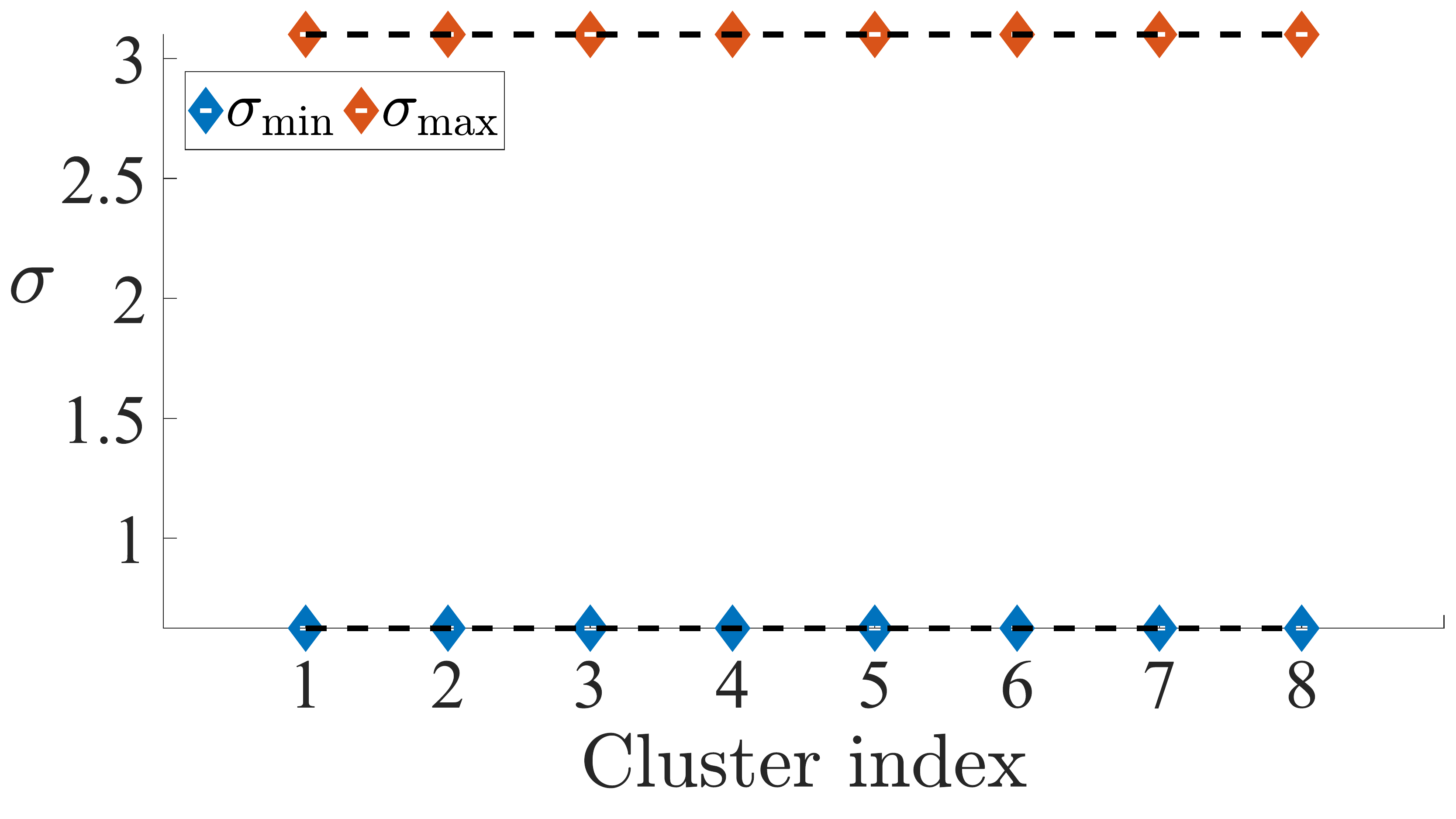}}
    \subfigure[]{\includegraphics[width=0.49\linewidth]{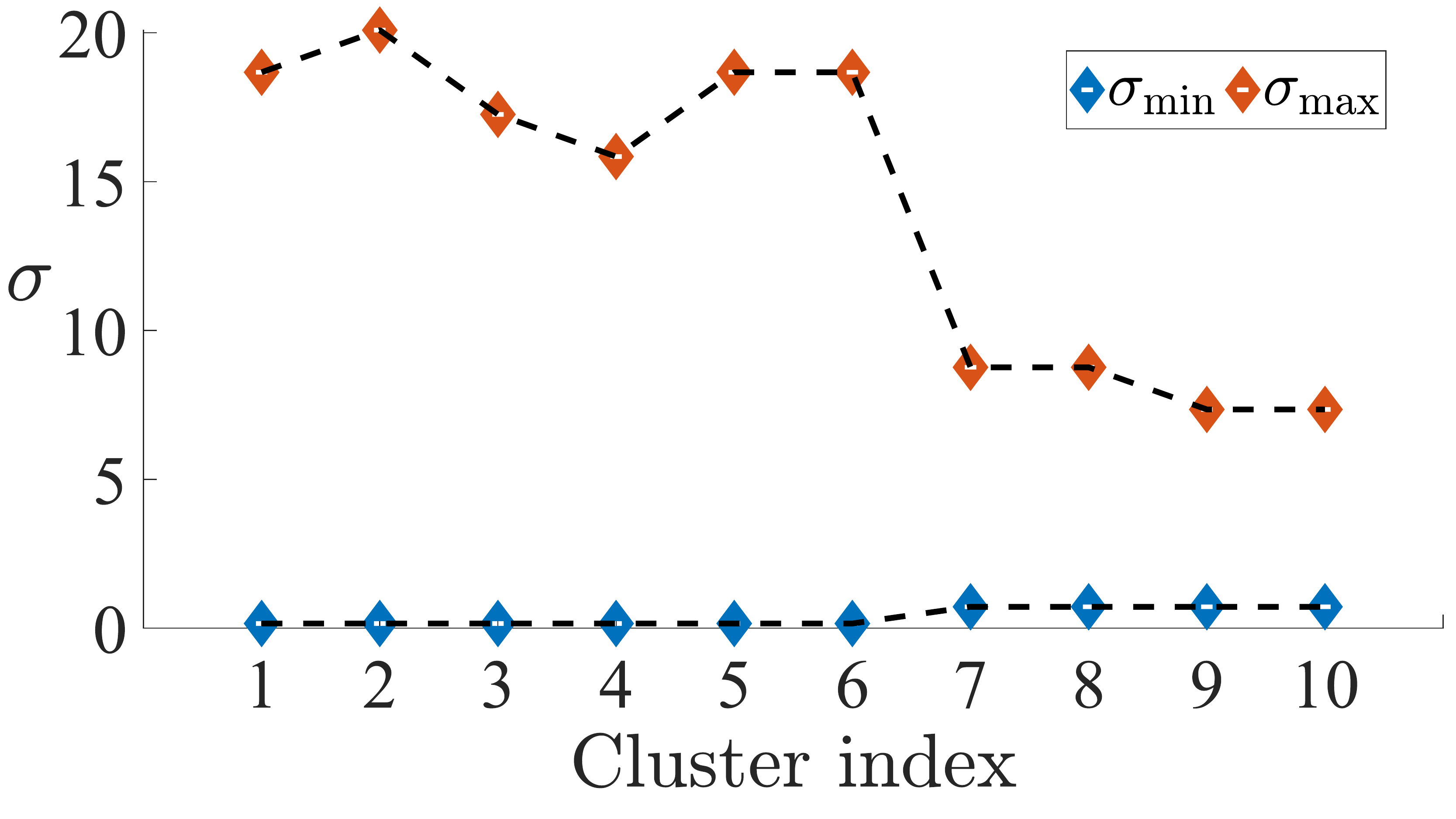}}
    \subfigure[]{\includegraphics[width=0.49\linewidth]{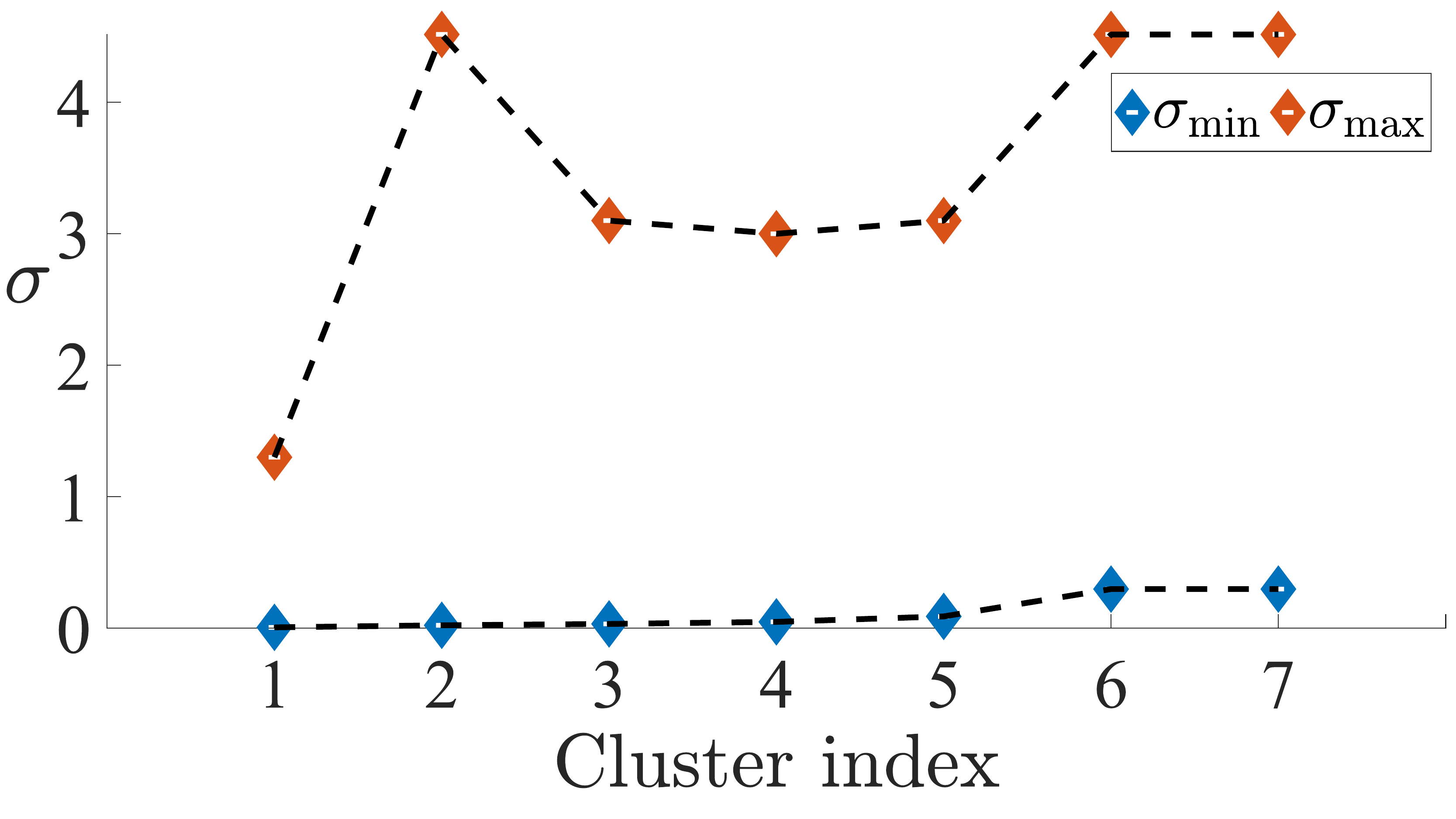}}
    \subfigure[]{\includegraphics[width=0.49\linewidth]{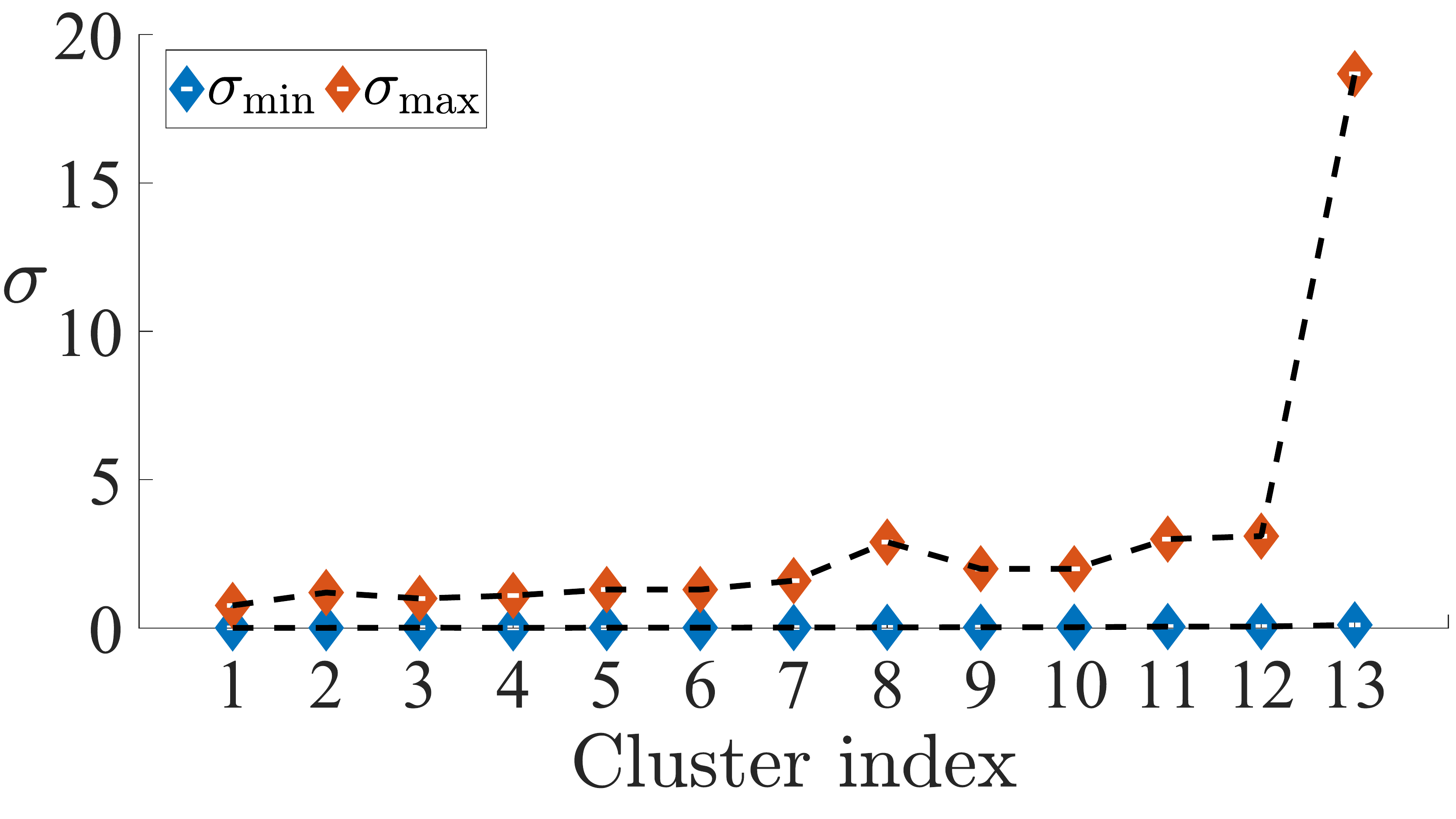}}
    \subfigure[]{\includegraphics[width=0.49\linewidth]{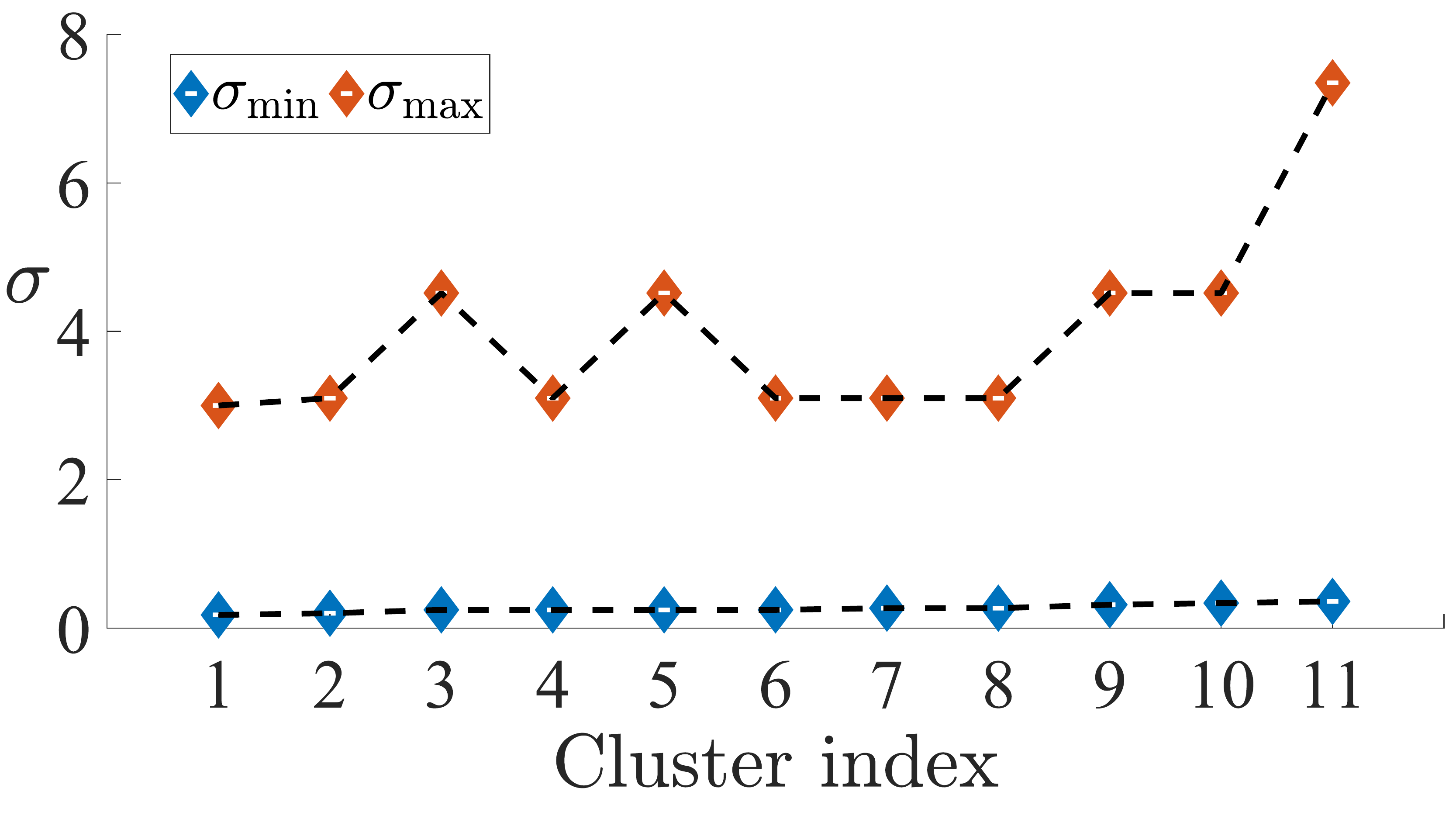}}
    \caption{Upper and lower bounds  $\sigma_{\max}$ and $\sigma_{\min}$, vs the cluster index for the five real networks in Table I. We set $\kappa$ such that $\mu^R_{\min}$ is the same and equal to $0.1$ for all the networks. Cluster indices refer to non-trivial clusters \textcolor{black}{and are in the order of increasing $\sigma_{\min}^k$, $k = 1, \hdots, \tilde K$}. Each panel shows a different network: (a) Power grid 1, (b) Power grid 2, (c) Neural network, (d) Brain network, and (e) Twitter network.} \label{fig:real1}
\end{figure}

 \textcolor{black}{Each panel of Fig. \ref{fig:real1} corresponds to one of the five real networks and shows the lower and upper bounds, $\sigma_{\min}^k$ and $\sigma_{\max}^k$, vs the cluster index $k$, after setting $\kappa$ such that for all these networks $\mu^R_{\min} = 0.1$. We see that for all the networks (except for Power grid 1 and Power grid 2) there is at least a pair of blocks that are related by Partially Disjoint CS. Clusters in Power grid 1 are all intertwined, and the blocks of Power grid 2 network are related by Matryoshka CS. A further analysis of the cluster synchronizability of these networks is provided in Sec.\ V of the SI, where the effect of varying $\kappa$ is investigated.}

\begin{table}[H]
\centering

\caption{\textbf{Real Networks}. For each network we report the number of nodes $N$, the number of edges $S$, and the minimum of the negative of the eigenvalues $\mu_{\min}^R$ (for $\kappa=0$.)  All networks are undirected and unweighted.}\label{table:real}

\begin{tabular}{l c c c c c}
\hline \hline
     Type & Name & $N$\,\, & $S$\,\, & $\mu_{\min}^R$\,\, & {\# of non-trivial clusters} \\
     \hline
     Power grid 1 & case16ci & 16 & 16 & -1.93 & 8 \\
     Power grid 2 & case85 & 85 & 84 & -1.62 & 10 \\
     Neural network & backward
circuit  & 29 & 59 & -1 & 7\\
     Brain network & macaque-rhesus-brain-2\, & 91 & 582 & 0 & 13 \\
     Twitter network & rt-retweet& 96 & 117 & 0 & 11 \\
     \hline \hline
     \end{tabular}
\end{table}


\section{Conclusion}

In this paper we introduced  definitions of synchronizability for cluster synchronization, which generalize the concept of synchronizability originally introduced for complete synchronization in \cite{barahona:2002}. We find that the case of cluster synchronization is by far more complex than the case of complete synchronization. We first considered cluster synchronization in networks with no intertwined clusters, \textcolor{black}{for which we study two problems: (i) introducing a definition of synchronizability for each cluster and (ii) introducing a definition of synchronizability for the entire network. For problem (i) we were able to derive a  dynamics-independent metric of synchronizability,  but for problem (ii) we were not.}

Our proposed definition of cluster synchronizability for the entire network is the range of the coupling strength $\sigma$ over which all the clusters synchronize. Characterizing this index for both the bounded case and the unbounded case requires considerations that generally depend on both the network structure and the dynamics.  \textcolor{black}{Hence, a main limitation compared to the case of complete synchronization is that our definition of cluster synchronizability for the entire network (Eq.\ \eqref{eta:General}) does not solely depend on the network topology, but also on the cluster indices $k_1$ and $k_2$, obtained from Eq.\ \eqref{GenEqCSynchclust} and so on the MSF bounds (see e.g., Fig.\ 3.)}
This motivated us to introduce and characterize three different types of cluster synchronization: Matryoshka CS, Partially Disjoint CS, and Complete Disjoint CS. Of these, to the best of our knowledge, only cases of Matryoshka synchronization had been previously reported (see Fig.\ 3 from \cite{pecora:2014cluster}). However, a study of several real networks from the literature shows that Partially Disjoint CS is common in these networks.

Finally, we also studied the general case of networks with intertwined clusters and performed an analysis of the cluster synchronizability for several real networks from the literature. 
Our work can be extended to the case of CS in multilayer networks \cite{della2020symmetries}.

\section*{Supplementary Material}
The supplementary material include further clarifications on the stability analysis of the synchronous solution for the cases of complete synchronization and cluster synchronization, an alternative formulation of the equations describing the network dynamics corresponding to the case $\kappa <0$, additional information on the canonical transformation that separates the quotient dynamics from the transverse dynamics, and an analysis of the synchronizability of several real networks as a function of the parameter $\kappa$.

\section*{Acknowledgement}
This work is supported by NIH grant 1R21EB028489-01A1.

\section*{Author Declarations}
The authors have no conflicts to disclose.

\section*{Data Availability}
The data that supports the findings of this study are available within the article.

 \newcommand{\noop}[1]{}

\end{document}